# Mechanisms of Microstructural Evolution and Degradation in Aluminum under High-Damage Irradiation

Alhassan S. Issaka, Sadie Wicks, Vishal Yadav, Assel Aitkaliyeva, Michael R. Tonks and Simon R. Phillpot*

*Department of Materials Science and Engineering, University of Florida, Gainesville, FL, 32611, USA*

**Abstract**

Aluminum alloys are widely used in research reactor systems, yet the mechanisms governing irradiation-induced degradation remain poorly understood. Here we combine conventional cascade-overlap molecular dynamics simulations with an accelerated Iterative Kinetic Approach (IKA) to investigate defect evolution in single-crystal Al subjected to 50 keV He irradiation. Benchmarking shows that IKA reproduces the essential defect kinetics of cascade simulations while enabling access to substantially higher accumulated damage. By extending the IKA to higher damage levels, we identified three distinct regimes governing radiation-induced degradation in Al: recombination-driven annihilation, defect accumulation, and sink-controlled absorption. At higher damage, Frank loops dissociate into Shockley partials and stair-rod loops, ultimately driving the nucleation and growth of stacking-fault tetrahedra (SFTs). These transformations progressively convert mobile defects into SFTs. Ultimately, the synergistic effect of interstitial and vacancy loops, and SFT increases irradiation hardening in Al at 300K. This work provides insight into irradiation-induced degradation in aluminum reactor materials.

* Corresponding author: phillpot@ufl.edu

## Introduction

High-energy atomic collisions in structural materials underpin many advanced material technologies, including semiconductor processing [1,2], aerospace manufacturing [3], and nuclear power generation [4]. In nuclear reactors, structural components and containment systems experience sustained degradation from prolonged exposure to energetic particles in harsh operating environments. Among materials used in nuclear applications, Al and its alloys are particularly important in research reactors because of their low neutron absorption cross-section, high thermal conductivity, and excellent corrosion resistance [5]. Al alloys are therefore widely used in fuel cladding, guide tubes, and reactor containers [6–8]. Notably, AA6061-T6 and AlFeNi alloys, containing approximately 97–98 wt.% and 60–90 wt.% Al, respectively, are extensively employed as core structural and fuel cladding materials in materials testing reactors, TRIGA reactors, and miniature neutron source reactors [5,9].

In Al alloys, irradiation significantly modifies the elastic-plastic behavior through the generation and evolution of radiation-induced defects. Irradiation first produces point defects, including vacancies and interstitial dumbbells with multiple orientations [10,11], which subsequently migrate and coalesce into three-dimensional defect clusters and prismatic dislocation loops [12]. Zhu *et al.* [13] observed dot-like features in Kr-ion-irradiated Ti-Al alloys, later identified as interstitial clusters. Similarly, irradiation of high-entropy fiber-strengthened Al-based alloys leads to the preferential nucleation of voids and dislocation loops as the dominant defect structures [14]. Under long-range diffusion conditions, vacancy clusters may collapse into three-dimensional stacking fault tetrahedra (SFTs), acting as strong obstacles to dislocation motion in Al [15]. At higher damage (~50 dpa) irradiation-induced precipitation has been reported in Al alloys, accompanied by the formation of large, densely distributed voids that are more pronounced than in non-Al systems [16].

While experimental studies of radiation damage in Al alloys are important, modeling and simulation can provide details of defect generation that cannot be obtained with experiments. Molecular dynamics (MD) simulations are widely used to investigate radiation damage evolution at the atomic scale. Radiation damage in MD can be broadly divided into two stages based on the timeline for production of defects: the ballistic and kinetic stages. The ballistic stage lasts approximately 1 ps and captures the initial formation of very rapidly moving point defects, characterized by intense thermal spikes. This is followed by the long-term kinetic stage, during which athermal and/or thermal recombination, defect clustering, and diffusion-driven processes dominate and ultimately govern the radiation response of the material [4]. Explicit MD simulation of the ballistic stage is computationally expensive due to the extremely rapid system evolution, which necessitates very small time steps. Moreover, coupling high-energy collision potentials such as the Ziegler–Biersack–Littmark (ZBL) screened nuclear repulsion can lead to over- or under-prediction of the threshold displacement energy. This affects the residual defects, owing to the strong sensitivity of cascade dynamics to the $r_1$–$r_2$ switching interval [17]. Despite this, the MD method serves as a useful tool for understanding defect formation due to radiation damage in Al.

Several studies have investigated the impact of radiation damage in Al using MD. Goryaeva *et al.* [18] investigated cumulative irradiation damage in Al up to 0.014 dpa and reported the nucleation of A15 nanophase clusters that precede dislocation loop formation. Da Fonseca et al. [19] examined the mechanisms governing irradiation creep in Al and showed that, at 0.03 dpa, applied stress promotes the preferential nucleation of well-oriented dislocation loops, leading to microstructural anisotropy. Their results demonstrate that stress-induced preferential nucleation dominates over stress-induced preferential absorption under electron irradiation. Using 1 MeV Kr ions through ~100 nm titanium aluminum alloy, Zhu *et al.* [13] observed that at ~ 1.57 dpa, local stress accumulation near large interstitial clusters and lamellar interfaces triggers the nucleation and propagation of stacking faults. Similarly, Jourdan et al. [20] reported the formation of three-dimensional compact A15 Frank–Kasper nanophases and their subsequent transformation into two-dimensional Frank loops at 0.03 dpa. Wang *et al.* [21] further showed that increasing the Al content in $Al_xCoCrFeNi$ alloys enhances radiation damage by increasing the peak Frenkel pair production due to a reduced threshold displacement energy and by prolonging the thermal spike at different PKA energies. Nevertheless, current existing MD studies of radiation damage in Al are limited to low-damage regimes because of the high computational cost of both the very short time steps required during the ballistic phase and the need to run thousands of cascade-overlap simulations. As a result, a comprehensive understanding of the mechanisms governing the nucleation and transformation of point defects into extended defects such as stacking fault tetrahedra and dislocation loops, remains lacking. Additionally, understanding how irradiation-induced microstructure affects the mechanical response in Al is limited.

In this study, we apply a cascade-informed accelerated method, the Iterative Kinetic Approach (IKA), to investigate irradiation damage evolution in aluminum under 50 keV $He^{2+}$ irradiation. We first benchmark the accuracy of IKA against conventional cascade overlap (CA) simulations using primary knock-on atom (PKA) events and then extend its application to higher damage levels. Through this framework, we systematically analyze the formation of point defects, their evolution into clusters, and the emergence of complex dislocation networks and stacking fault tetrahedra (SFTs), in qualitative agreement with experimentally observed irradiation-induced SFTs in Al and Au [15,22]. We further provide mechanistic insights into dislocation faulting and unfaulting reactions and elucidate the role of intrinsic irradiation-induced stress on defect dynamics. Finally, we quantify the contributions of different defect obstacles to irradiation hardening in Al. This work establishes a mechanistic link between radiation-induced defect evolution and the resulting mechanical response, providing insight into microstructural degradation in Al reactor materials

The remainder of the paper is organized as follows. First, we describe the simulation methodology, including cascade-overlap simulations, the IKA method, and defect analysis procedures. next, we validate IKA by benchmarking its predictions against conventional cascade-overlap simulations, followed by an exploration of high-damage defect evolution, including isolated defect kinetics, vacancy clustering, interstitial clustering, dislocation formation, SFT nucleation and irradiation-induced hardening. Finally, we discuss these results in the context of existing literature and provide our conclusions.

## Results

This section is divided into two main parts. First, we benchmark the evolution of defects (point defects and clusters) from the IKA against full conventional cascade-overlap simulations and then we extend the IKA to higher damage levels and characterize the resulting defect evolution.

### Dynamics of Defects: IKA versus Cascade overlap

Here we present results from consecutive cascade-overlap approach (CA) and consecutive IKA simulations to understand how the dynamics of isolated point defects and cluster evolutions compare between the two methods. Fig. 1a compares the total defect fraction, defined as the sum of isolated point defects and atoms in all clusters divided by the total number of atoms in the simulation cell (276,480).

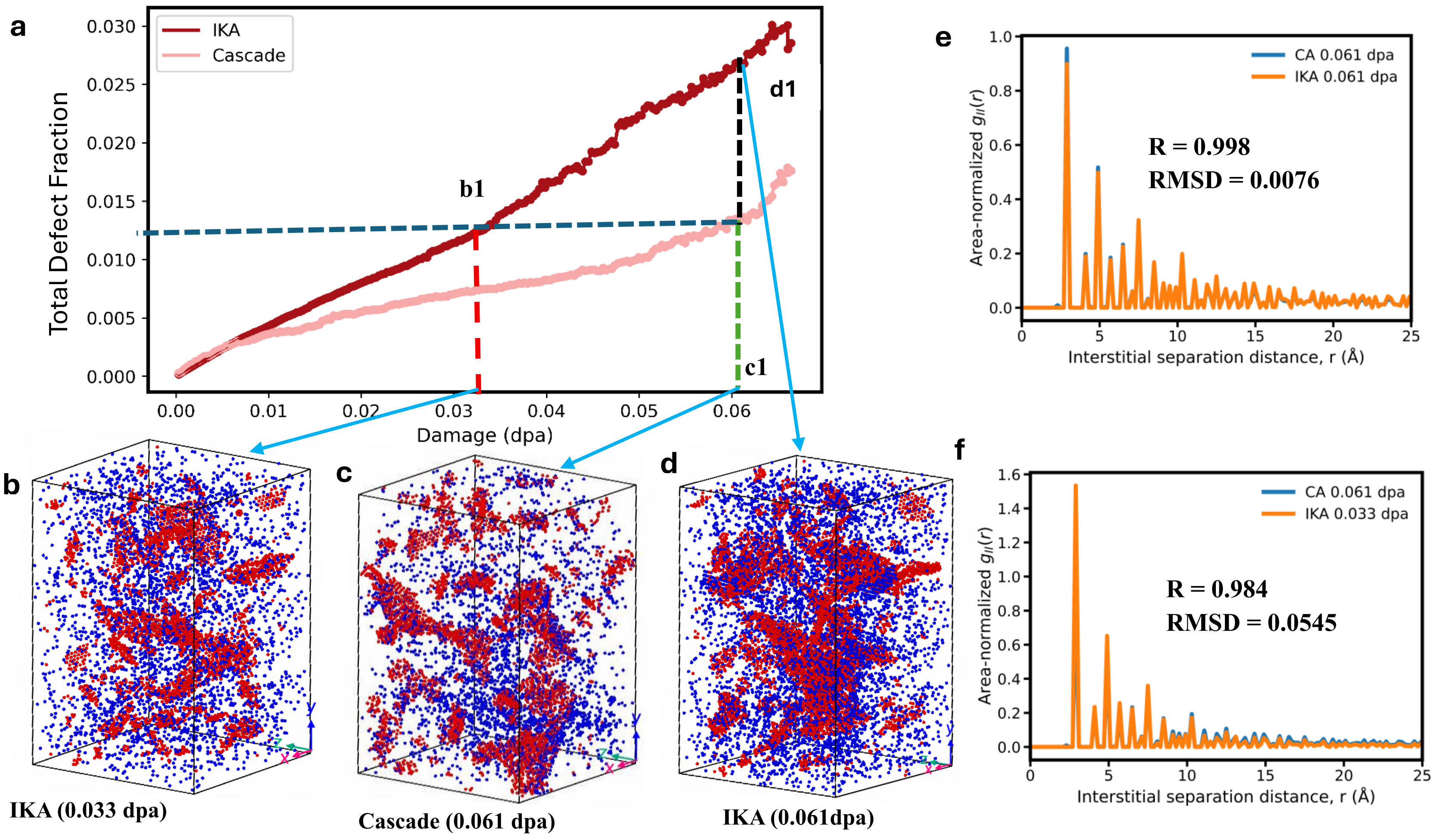

*Fig. 1. Comparison of defect evolution and microstructure predicted by IKA and CA simulations. a) Total defect fraction (off-lattice atoms) as a function of damage. The red dashed line b1) marks the IKA damage (0.033 dpa) at which the total defect fraction matches that of the CA simulation at 0.061 dpa c1), while the black dashed line d1) denotes the IKA state at the same nominal damage of 0.061 dpa. Representative atomic snapshots of IKA at b) 0.033 dpa, c) CA at 0.061 dpa, and d) IKA at 0.061 dpa. e) Area-normalized interstitial pair-correlation comparison between CA and IKA at the same damage (0.061 dpa). f) Corresponding comparison between CA at 0.061 dpa and IKA at 0.033 dpa, selected to match the total defect fraction. Red atoms are interstitials and blue atoms are vacancies*

At early stages of damage, both methods exhibit nearly identical defect fractions. As damage accumulates, however, the IKA produces defects more rapidly than the CA, such that by 0.061 dpa the total defect fraction in the IKA is approximately twice that of the CA. Notably, the defect fraction obtained from the IKA at 0.033 dpa is comparable to that from the CA at 0.061 dpa. Representative three-dimensional snapshots from the IKA at 0.033 dpa, and the CA at 0.061 dpa are shown in Fig. 1b and 1c, respectively, and reveal strikingly similar defects densities. Comparison of CA and IKA at 0.061 dpa (Fig. 1c, d) shows that IKA predicts approximately twice the number of stable defects relative to CA. To determine whether this difference in defect concentration affects the local defect arrangement, we performed area-normalized interstitial–interstitial pair-correlation analysis for two cases: (i) same damage (dpa) comparison between CA 0.061 dpa and IKA 0.061 dpa (Fig. 1e), and (ii) an equivalent-defect-fraction comparison between CA 0.061 dpa and IKA 0.033 dpa (Fig. 1f). In both cases, the dominant peaks occur at the same interstitial separation distances and evolve with overlapping peaks, indicating that the spatial arrangement of interstitials is preserved. At same damage (0.061 dpa), the two curves are nearly indistinguishable, with a Pearson correlation coefficient of R=0.998 and an RMSD of 0.0076. When compared at equivalent defect fraction (CA 0.061 dpa vs. IKA 0.033 dpa), the agreement remains strong, with R=0.984 and RMSD = 0.0545, although the larger RMSD reflects a greater concentration of pair probability at short interstitial separations in the lower-damage IKA configuration than CA at 0.061 dpa. These results indicate that, despite higher number of surviving defects, IKA reproduces the same characteristic interstitial spacing and local defect topology observed in the cascade-overlap simulations. Because defect clustering kinetics are strongly damage/dose dependent, agreement at a single damage level is insufficient to establish methodological equivalence, thus comparing IKA and CA simulations across a range of damage levels enables assessment of method fidelity and mechanistic equivalence. The details for comparing different damage levels for the two methods is discussed next.

Results of interstitial and vacancy point defects for both methods are shown in Fig. 2. In Fig. 2a, the fraction of isolated interstitials from the consecutive simulations in both methods drops significantly, falling from ~1 at the lowest damage to a shallow plateau of 0.06 – 0.10 by 0.02 dpa. However, we notice a small offset between 0.005 dpa and 0.03 dpa where the fraction of isolated interstitials from the CA is higher than the IKA. This implies that between 0.005 – 0.03 dpa, the IKA exhibited a higher interstitial clustering than the CA. We attribute such slight offset in the CA method to the highly energetic environment in which defects are produced in cascades. The ballistic/thermal-spike dynamics can inhibit early interstitial clustering by disrupting weakly bound embryonic aggregates, leaving a slightly larger fraction of interstitials as isolated defects in the 0.005 – 0.03 dpa regime.

Beyond 0.03 dpa however, the fraction of isolated interstitials in both converges and remains the same until the end of the simulation. Such close overlap of the trends indicates that the two methods follow the same kinetics of interstitial evolution where isolated interstitials rapidly cluster or recombine with other defects, after which a quasi-steady state is established. As shown in Fig. 2b, the fraction of isolated vacancies exhibits a decreasing trend with increasing damage. In contrast to the isolated interstitial behavior in Fig. 2a, the IKA consistently retains a higher fraction of isolated vacancies than the CA, indicating reduced vacancy clustering under the same conditions as the CA. Unlike interstitials, which rapidly cluster, vacancies manifest much slower clustering kinetics.

While the absolute number of clusters (Fig. S3a & S3b) highlights how efficiently isolated point defects in each method are captured into clusters, it does not tell us how large those clusters are or the coarsening/growth behavior. To understand this, we show the average interstitial cluster sizes in Fig. 2c. The IKA shows a sharp, nearly linear increase in average interstitial cluster size, reaching values above 12 atoms per cluster by 0.02 dpa while the average cluster size in the CA is only 6 atoms per cluster at 0.02 dpa. At 0.035 dpa the average cluster size from the IKA peaks with 19 atoms per cluster, while average cluster size from the CA is 11 atoms per cluster. The CA takes more damage to reach peak cluster size, reaching it around 0.05 dpa with 13 atoms per cluster. Although CA produces more interstitial and vacancy clusters at lower damage (0.005 – 0.03 dpa, Fig. S3a & S3b), these clusters remain smaller on average than in the IKA. However, the average cluster size analysis shows that they both converge at 0.048 dpa (Fig. 2c) with similar average cluster size thereafter. This suggests that differences in cluster coarsening between IKA and CA only exist at lower damage. Beyond 0.048 dpa threshold, they converge and proceed with similar kinetics. Next, we examine the evolution of vacancy clusters in the IKA and CA simulations, shown in Fig. 2d. Although the average cluster size decreases monotonically with accumulated damage in both cases, the IKA consistently yields slightly smaller vacancy clusters than the CA, with the difference progressively diminishing as damage increases. Like the interstitial case, the vacancy cluster sizes from the two methods converge to nearly the same value at ~0.048 dpa.

We now examine the microscopic characteristics of interstitial clustering in the IKA and CA simulations. Fig. 3a–d shows the distributions of interstitial cluster counts across cluster-size ranges at four damage levels (We show the full cluster size distribution in the supplementary material, Fig. S4, a-i). At all damage levels, the distributions are strongly left-skewed, with clusters containing 2–4 and 5–7 interstitials dominating the population and intermediate-sized clusters (8–10 to 21–25 interstitials) forming a modest shoulder. With increasing damage, both methods show a coherent evolution of the full spectrum: while small clusters remain most abundant, intermediate-sized clusters become increasingly populated, and the large-cluster tail grows, reflecting continued cluster growth and coalescence.

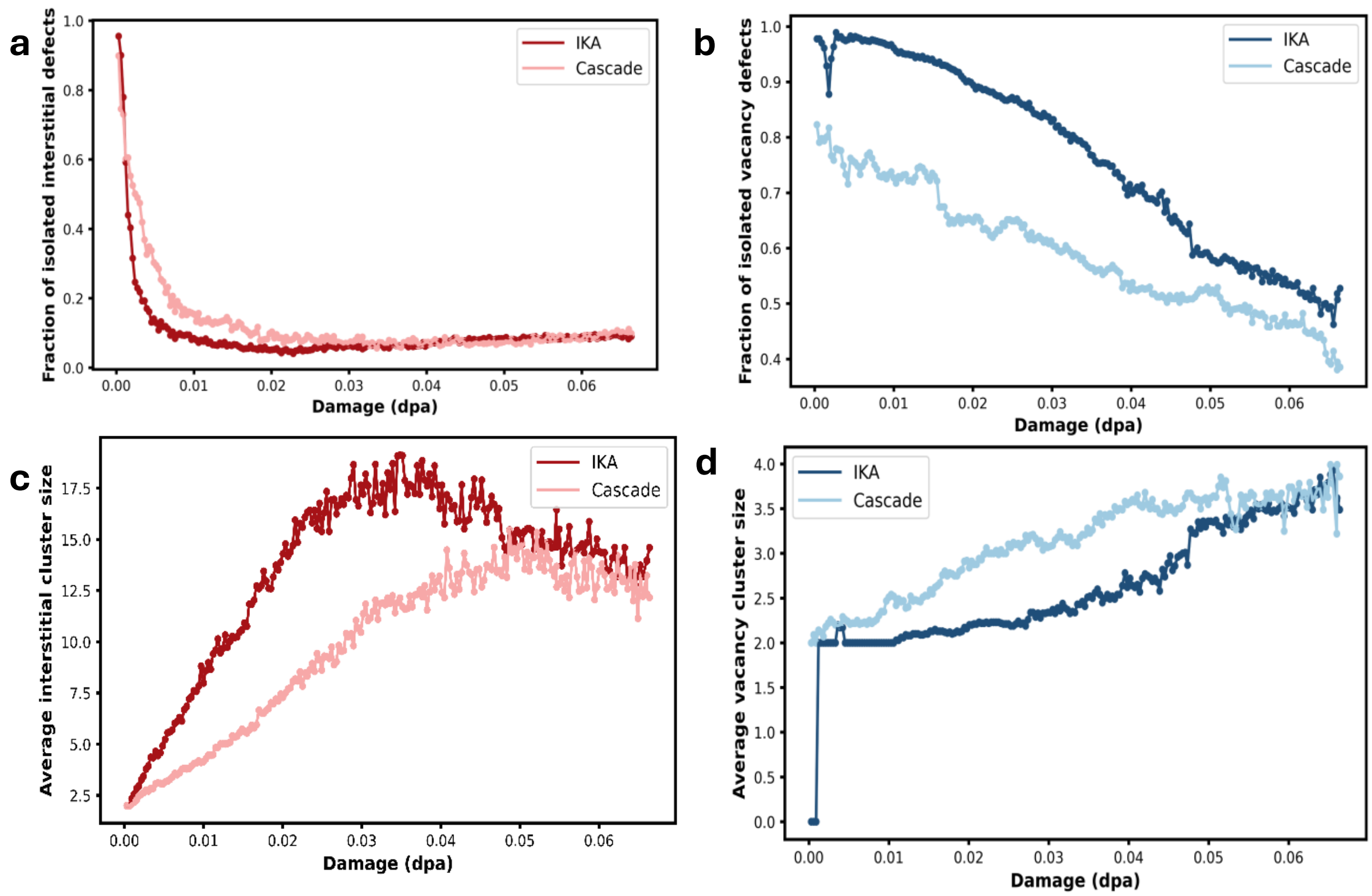

*Fig. 2. Comparison of interstitial and vacancy defect evolution predicted by IKA and CA as a function of damage. a) Fraction of interstitials that are isolated. b) Fraction of vacancies that are isolated. Plots a and b are normalized to the total defects, c) Average interstitial cluster size (number of interstitials per cluster) and d) Average vacancy cluster size (number of vacancies per cluster).*

At 0.0015 dpa (Fig. 3a), the CA produces marginally smaller clusters than the IKA; at 0.0305 and 0.0482 dpa (Fig. 3b-c) the distributions between the two are largely similar. With 0.0581 dpa (Fig. 3d), the highest damage examined, the two distributions remain broadly similar, with slight differences confined primarily to the smallest and largest cluster sizes. Further evidence of such close similarity after 0.03 dpa is the reduced heterogeneity between the ratio of the smallest and largest cluster sizes shown in Fig. S5. To understand the evolution of interstitial clustering with increasing damage, we compute the fraction of interstitials in each cluster-size range, as shown in Fig. 3e. As the damage increases, the interstitials redistribute from small clusters of 2 – 10 toward large clusters. Importantly, IKA and the CA exhibit very similar atom populations and evolution, with only a modest early-damage offset in the onset of very large clusters, indicating both approximate the same physical coalescence. Overall, the formation of point defects, their clustering behavior, and resulting defect distributions show close agreement between the IKA and CA methods. Therefore, we now proceed with only IKA to study higher damage effects.

**Total Defect Concentration at High Damage**

Having established that the IKA predicts similar defect behavior to the CA, we now apply it to go to three times as much damage, which would be computationally very burdensome with the CA. To understand the impact of cumulative defect buildup, we examine the total defect fraction defined as the sum of isolated point defects and atoms in all clusters divided by the total number of atoms in the simulation cell (276,480) with increasing damage in Fig. 4. This is an extension of Fig. 2 to higher damage. Because the simulation is performed on a single crystal, the total number of vacancies and interstitials are equal. At low damage levels, the defect fraction rises sharply, reaching nearly 5% at 0.1 dpa, before significant recombination or sink absorption occurs. Between 0.1 dpa and 0.15 dpa the fraction of total defects start decreasing. Beyond 0.15 dpa, a dynamic equilibrium is established in which defect formation is balanced by annihilation and sink absorption, leading to a slight decrease in the defect fraction.

During damage accumulation, majority of primary defects exist initially as isolated point defects, either single self-interstitial atoms (SIAs) or isolated (mono) vacancies, i.e., defects that are not part of clusters. Fig. 5 presents the evolution of defect annihilation, defect accumulation, the fraction of single SIAs and vacancies, and the concentration of isolated point defects up to 0.175 dpa. The cumulative fraction of annihilated defects, shown in Fig. 5a, is calculated as:

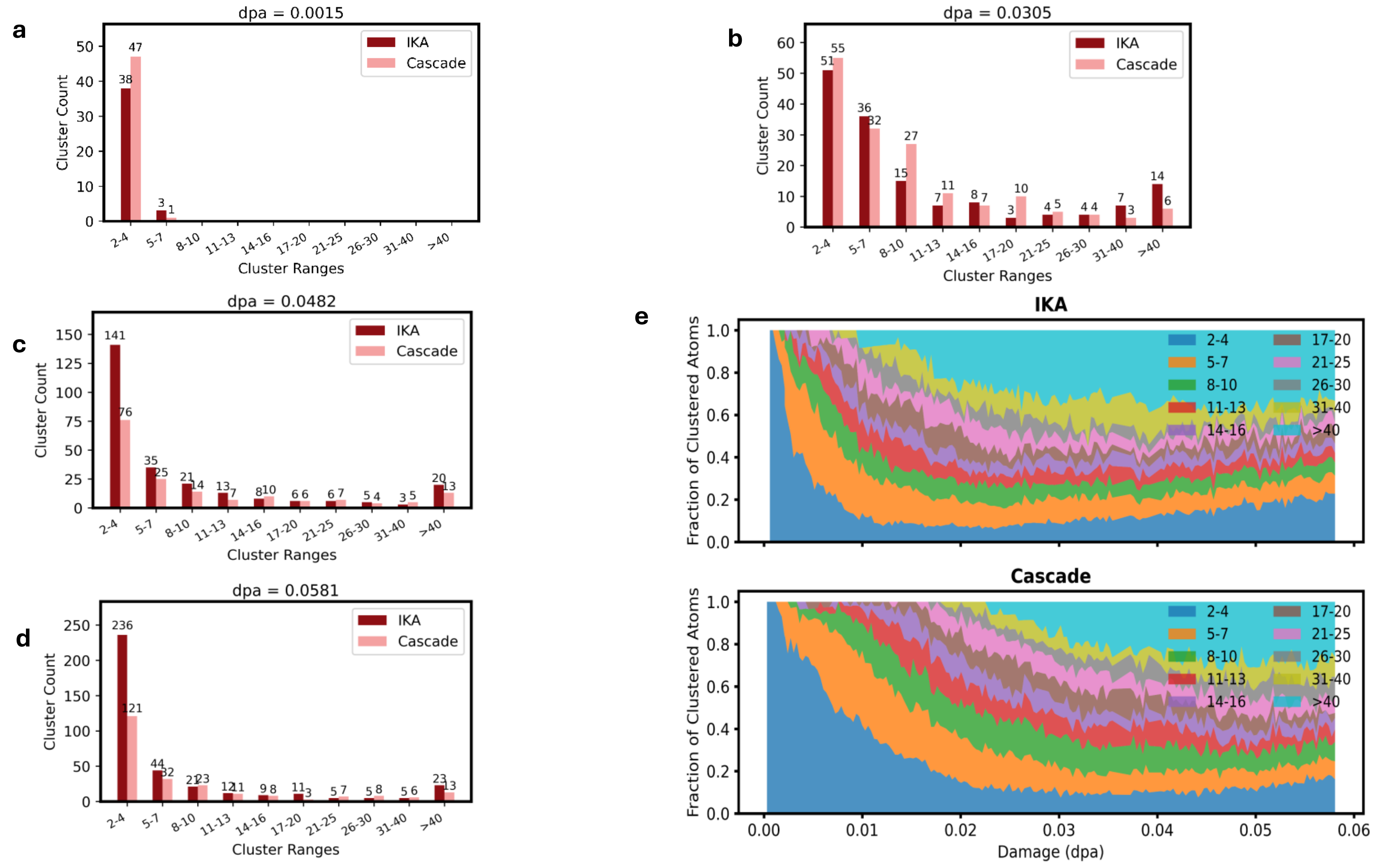

***Fig. 3.*** *Comparing interstitial cluster size distribution from IKA and CA. Comparing the distributions of the number of interstitial clusters of various sizes from the IKA and CA for increasing amounts of damage from a) 0.0015 dpa, b) 0.0305 dpa, c) 0.0482 dpa, d) 0.0581 dpa and e) shows the fraction of clustered interstitials in 10 different size ranges.*

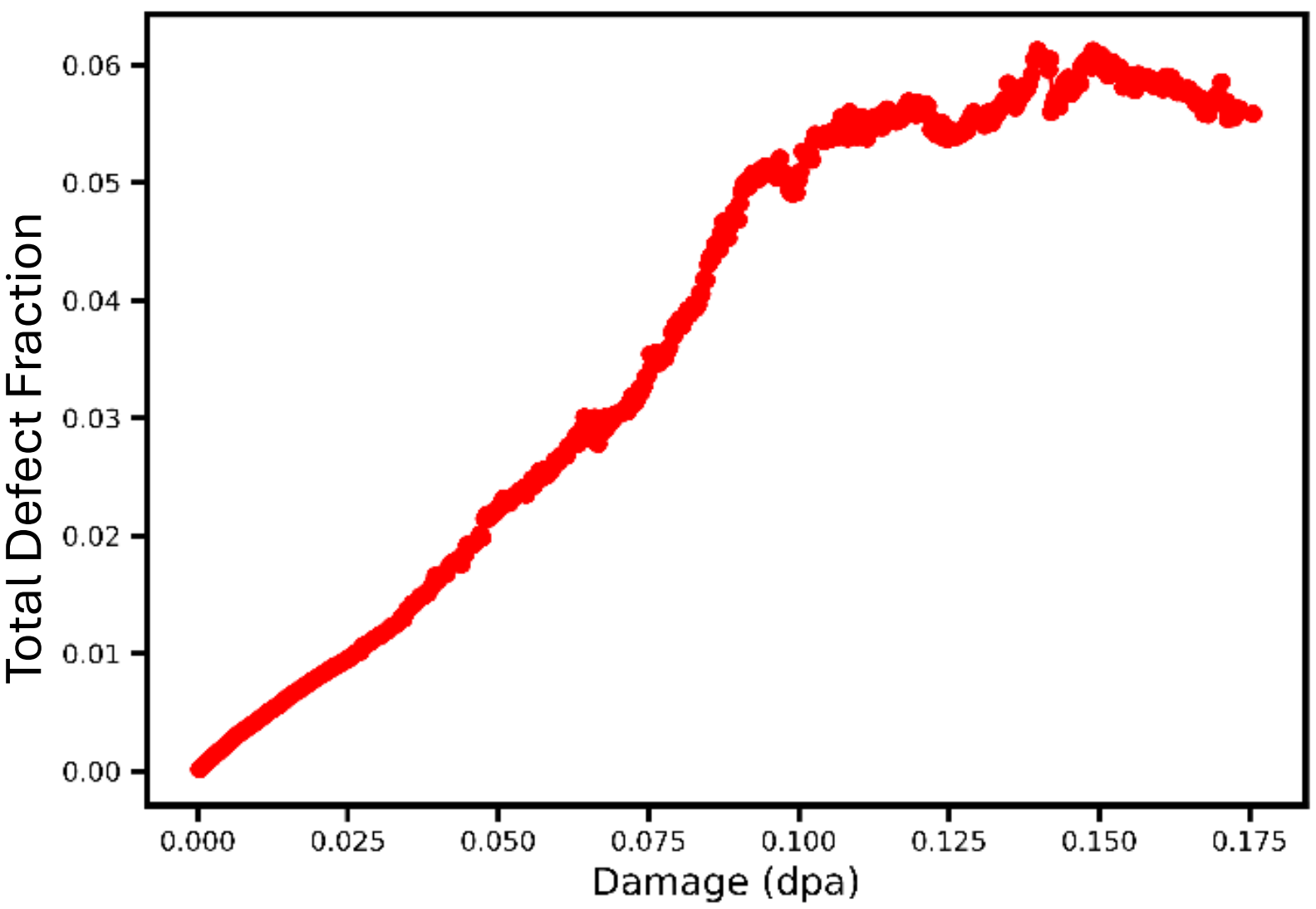

***Fig. 4.*** *Extension of the IKA defect concentration (fraction) evolution in Fig. 1a to higher damage. This is the defect concentration throughout the entire damage up to over 0.175 dpa, three times much damage than the CA.*

$$A_{cum}(d) = \left(1 - \frac{N_{surv}\,(d)}{N_{FP,Ins}(d)}\right) \tag{1}$$

where $A_{cum}(d)$ is the cumulative annihilation fraction at damage d, $N_{surv}\,(d)$ is the total number of surviving defects at damage d, $N_{FP,Ins}(d)$ is the total inserted FPs from the iterative insertions. Fig. 5a and 5b reveal three distinct regimes governing the entire defect evolution. In the first regime (Regime I), the annihilation fraction increases from nearly zero to ~ 29% at 0.03 dpa. This behavior is attributed to the high mobility of freshly created defects, particularly interstitials, which can migrate relatively freely due to the low defect density and limited obstacles. As a result, recombination between vacancies and SIAs is highly efficient. At this stage, absorption at sinks is minimal because the dislocation density is still low and insufficient to provide significant sink strength (See Fig. S1). Consistently, Fig. 5b shows that the defect survival (accumulation) fraction decreases from 1.0 to approximately 0.70 over the same damage interval, directly mirroring the increase in annihilation.

In the second regime (Regime II), the annihilation fraction decreases from ~29% to nearly zero at 0.09 dpa, while the defect survival fraction increases (Fig. 5b). This indicates that recombination becomes progressively less efficient as defect density increases. As damage accumulates, defect-defect interactions and clustering become more probable, limiting the ability of mobile interstitials to recombine with vacancies. Instead, defects increasingly participate in cluster formation, reducing the net recombination rate.

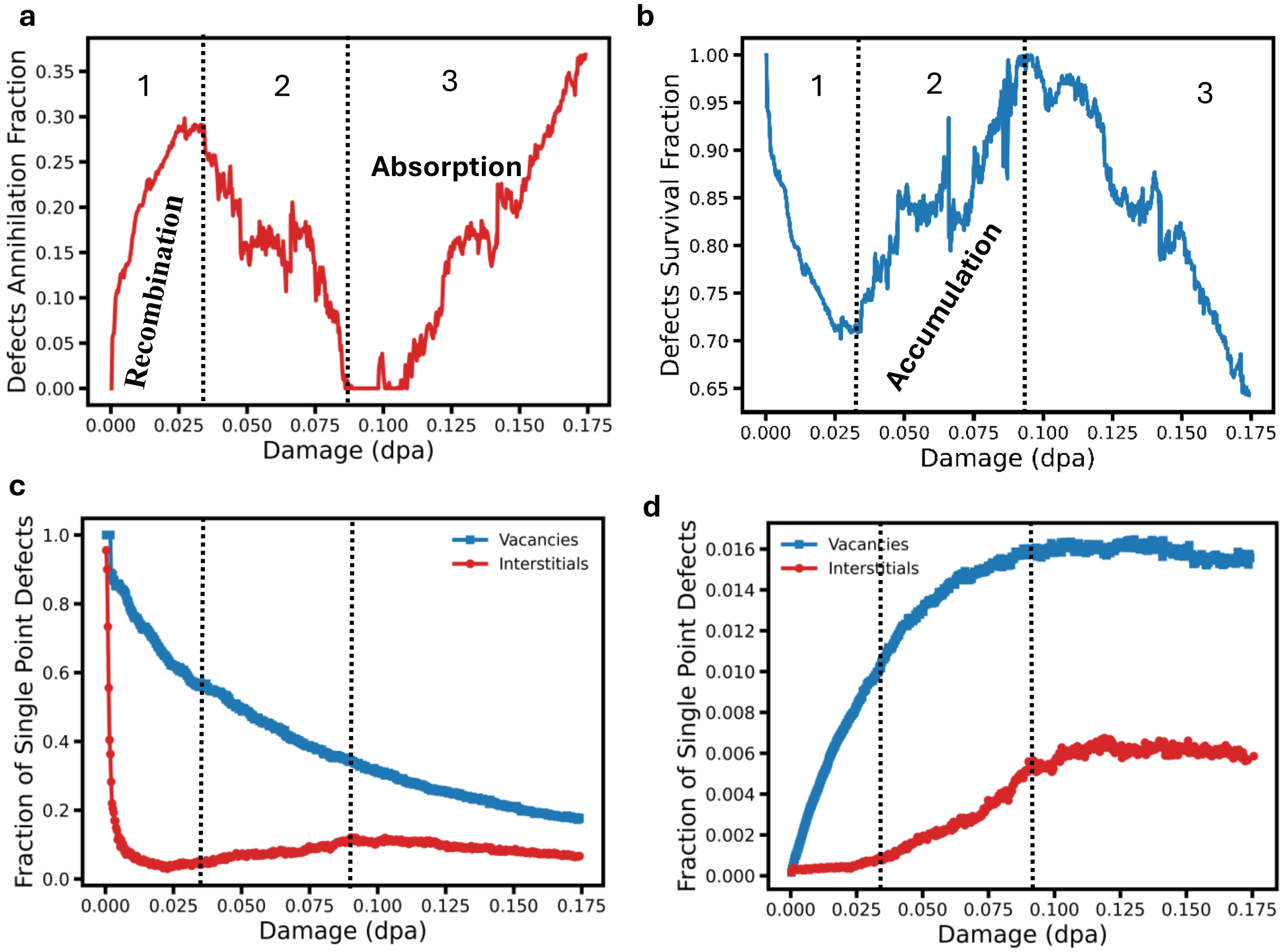

***Fig. 5***. *Regimes of different defects evolution up to 0.175 dpa. a) Fraction of all inserted defects annihilated showing the recombination and sink control absorption regimes, b) Fraction of all inserted defects that have survived/accumulated, thus depicting defects accumulation regime, c) Fraction of single point defects. a-c is normalized by the total inserted defects (FPs), d) concentration of single point defects (normalize by total atoms in the cell). The numbers 1, 2 and 3 represent the three damage regimes which are discussed in the discussion section.*

Beyond 0.09 dpa, the annihilation fraction begins to increase again, while the survival fraction decreases. This third regime corresponds to the activation of sink absorption mechanisms, primarily through the formation of dense dislocation networks and extended defect structures that act as efficient sinks. Thus, although recombination is suppressed in Regime II, annihilation re-emerges in Regime III due to enhanced sink absorption rather than direct vacancy–interstitial recombination.

Fig. 5c shows the fraction of single (non-clustered) point defects normalized by the total inserted Frenkel pairs. Both single interstitials and isolated vacancies exhibit a non-monotonic decrease with increasing damage, as expected. Notably, single interstitials decrease much more rapidly and stabilize by 0.025 dpa. This faster reduction reflects their lower migration barriers and higher mobility, which promote clustering or absorption at sinks. In contrast, isolated vacancies persist longer, consistent with their lower mobility. Fig. 5d presents the concentration of isolated SIAs and vacancies. The concentration of isolated vacancies increases rapidly with damage and reaches a plateau near 0.016 at approximately 0.075 dpa. By comparison, isolated SIAs rise more gradually and plateau around 0.006 at 0.125 dpa. Because Frenkel pair production creates equal numbers of vacancies and interstitials, the approximately threefold higher steady-state concentration of isolated vacancies indicates that vacancies are significantly less likely to cluster than SIAs. In other words, interstitials preferentially form clusters or are absorbed at sinks faster, while a larger fraction of vacancies remain isolated. The plateau behavior observed in Fig. 5d indicates the onset of a saturation regime in which defect generation is balanced by recombination and sink absorption processes. To further elucidate the physical origin of these three regimes, we next examine the evolution of defect clusters.

In Fig. 6a, the size of the average vacancy cluster increases very slowly. The average cluster size is around 2.5 atoms up to around 0.025 dpa, as shown in Fig. 6b. The total fraction of the atoms in vacancy clusters is very small during this time. Above 0.025 dpa the fraction of atoms in vacancy clusters increases quickly and the average cluster size also begins to grow. This increase coincides with the decrease in the recombination fraction shown in Fig. 5b. At 0.09 dpa, the rate of increase in the vacancy cluster fraction slows substantially and the average cluster size converges to about 5 vacancies. These changes coincide with the second increase in the recombination fraction.

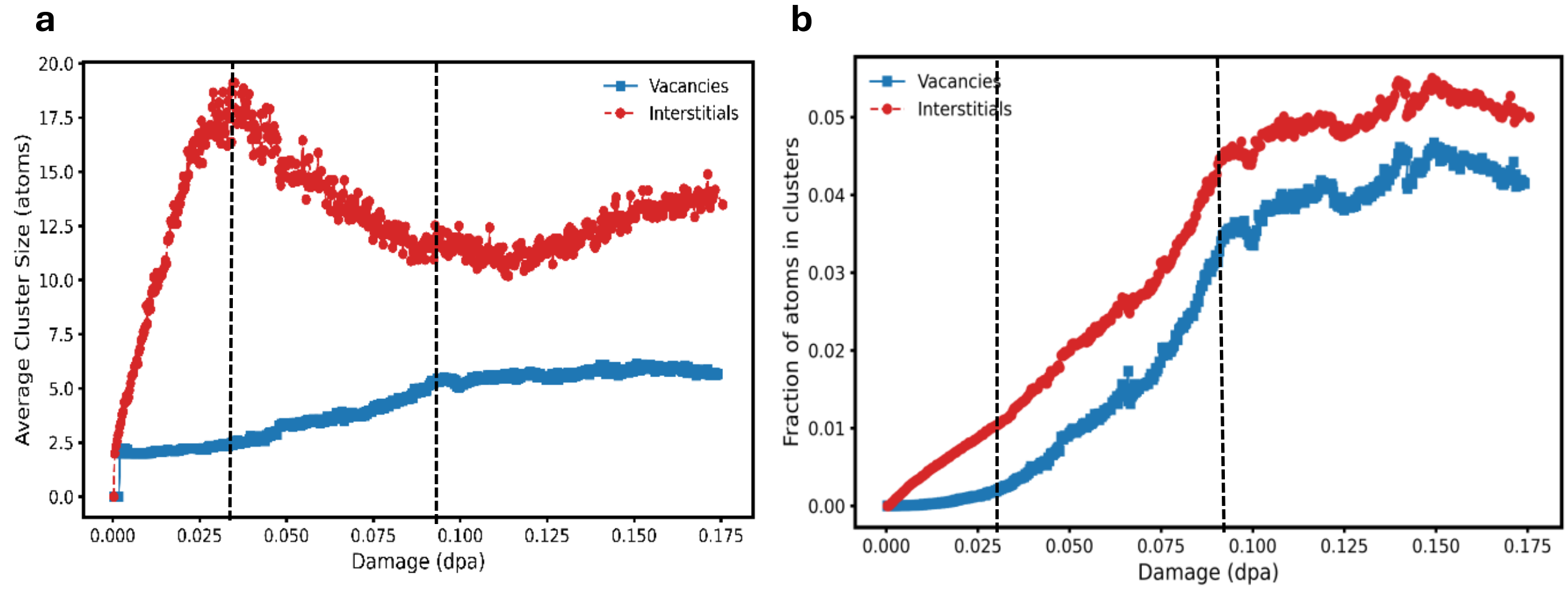

***Fig. 6.*** *Defect Clustering up to 0.175 dpa, a) Average cluster size as a function of increasing damage, b) fraction of all atoms involved in interstitial or vacancy clusters. The black dashed lines separate the three defect evolution regimes.*

Fig. 6a also presents the average interstitial cluster size as a function of damage and Fig. 6b the fraction of atoms in interstitial clusters. Unlike vacancy clusters, the average cluster size and the fraction of atoms in clusters increase linearly with time at the start of the simulation. The highly mobile interstitials quickly migrate and agglomerate, increasing the number and size of the interstitial clusters up to average size of 19 interstitials at 0.03 dpa, seven times larger than the vacancy clusters at this damage. Between 0.03 and 0.09 dpa, the interstitial cluster size decreases linearly, dropping to around 12 atoms by 0.09 dpa, although the fraction of atoms in interstitial clusters continues to increase during this period. This coincides with the decrease in the recombination fraction from Fig. 5b. Beyond 0.09 dpa, the average interstitial cluster size and interstitial cluster fraction change slowly for the rest of the simulation. In this regime, the average interstitial cluster size is twice the average vacancy cluster size, and the vacancy cluster fraction is around 20% lower than the interstitial cluster fraction. However, the vacancy cluster density is a factor of two higher than the vacancy interstitial clusters and levels off near 0.45 $nm^{-3}$ (Fig. S2).

Combining the information from Figs. 5 and 6, it appears that during the first regime which spans from the start of the simulation to 0.03 dpa, the isolated vacancy fraction in Fig. 5c decreases while the vacancy cluster size and vacancy cluster fraction do not change. In addition, the SIA fraction in Fig. 5c decreases substantially and the average interstitial cluster size and interstitial cluster fraction increase. This results in an increasing recombination fraction. During the second regime, spanning from 0.03 dpa to 0.09 dpa, the isolated vacancy fraction decreases slowly and the vacancy cluster size and fraction increase, while the SIA fraction decreases and interstitial cluster fraction increases but the average interstitial cluster size decreases. All this results in a decrease in the recombination fraction because enough clusters have formed to impede recombination. Finally, during the final regime, starting at 0.09 dpa, the fraction of isolated vacancies (Fig. 5c), SIA, vacancy clusters, and interstitial clusters remained steady, as is the average vacancy cluster size. The average interstitial cluster size decreases slightly and then increase gradually, and the absorption fraction (absorption because it is the dominant mechanism in this damage window) increases rapidly probably due to enough sink strength. Thus, increases in vacancy clustering correspond with higher absorption sink whilst low vacancy clustering corresponds to higher recombination regime.

It is also valuable to see the full distribution of cluster sizes. We first examine the distributions of interstitial clusters in the three regimes in Fig. 7a – c. Each of the plots shows the number of clusters within the specified cluster size range. At 0.0048 dpa (Regime I) in Fig. 7a, the cluster distribution is heavily skewed towards the smallest clusters where over 85 % of all clusters contain between 2 and 7 atoms, with only eight clusters in the 8 – 10 atom range and a negligible tail beyond 10 atoms. At this regime, the fast-moving interstitials are just beginning to interact and cluster. By 0.0430 dpa (Regime II) in Fig. 7b, the total number of clusters more than doubles, and the distribution broadens significantly. The number of 2 – 4 atom-sized clusters nearly triples and larger clusters form, including 18 with more than 41 interstitials. At 0.1068 dpa (Regime III) in Fig. 7c, the number of 2 – 4 sized clusters have again tripled, and all other sizes have at least doubled, except for the largest size which increased by 50%. The size of the clusters continues to skew toward small clusters because each IKA iteration seeds new FPs, reducing the relative fraction of larger clusters.

In Fig. 7d and e we show the vacancy cluster size distribution at 0.0430 and 0.1068 dpa; the results for 0.0048 dpa are not shown because very few vacancy clusters had formed. At 0.0430 dpa, vacancy clusters are small, in the 2–4 atom range, with very few larger clusters. At 0.1068 dpa, Fig. 7e, there is a significant growth in vacancy cluster size diversity, with at least four clusters in all cluster ranges. However, the vast majority of the clusters are still in the 2-4 atom range, with more than twice as many vacancy clusters in this range than interstitial clusters. Finally, in Fig. 7f, we estimate the ratio between the number of clusters in the largest and the smallest size bins (>41 atoms / 2–4 atoms) to get an idea of the degree of heterogeneity in clustering between vacancies and interstitials. It shows a clear asymmetry in clustering dynamics between the two. Interstitials exhibit a strong early peak in the large-to-small cluster ratio, reflecting rapid mobility-driven coalescence into larger clusters up to 0.03 dpa, followed by stabilization at higher damage. In contrast, vacancy clustering is strongly suppressed at low damage and increases only weakly at higher damage, consistent with diffusion-limited accumulation and delayed formation of larger vacancy clusters over the entire damage range. Overall, interstitials cluster more readily than vacancies.

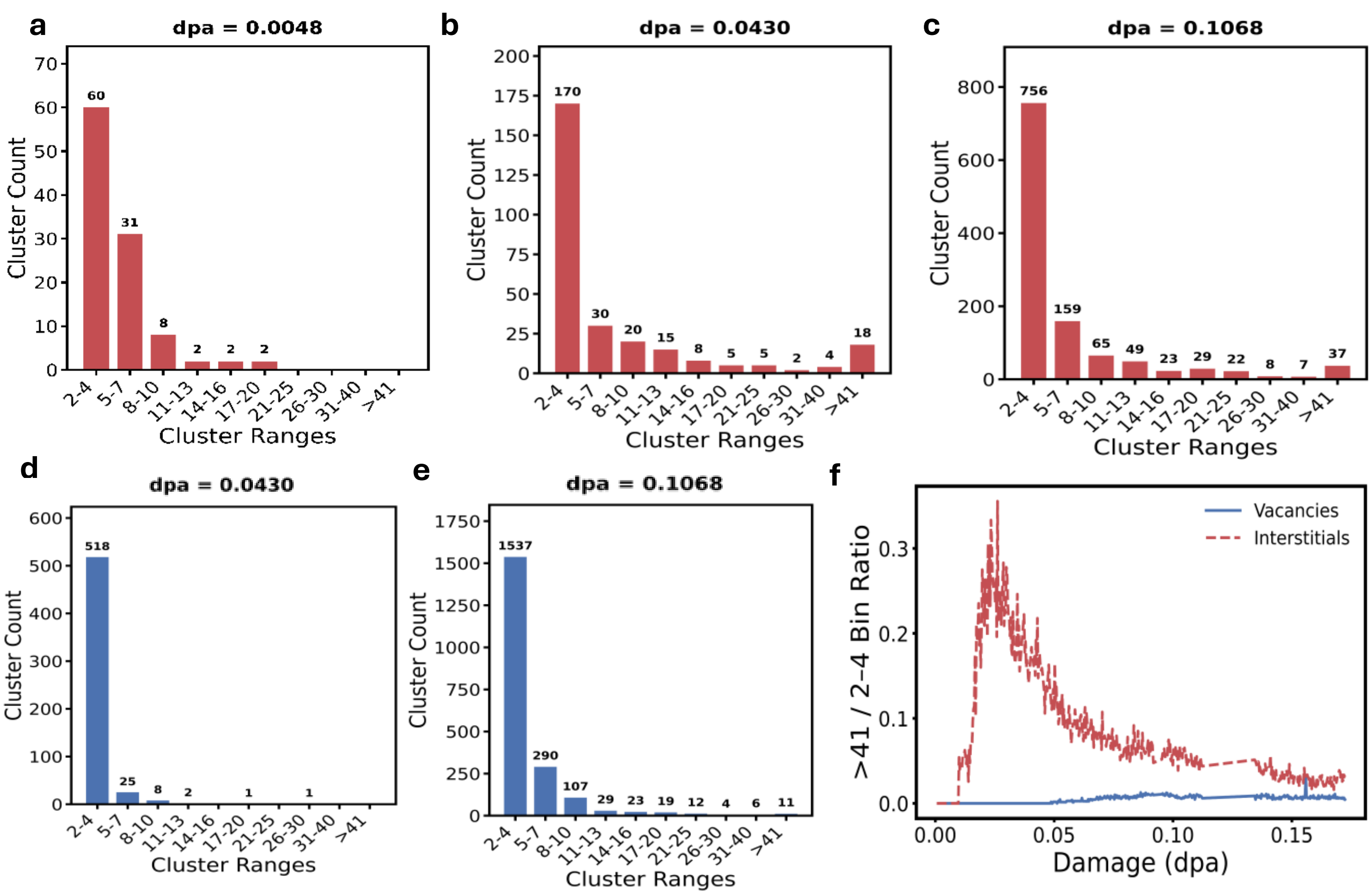

*Fig. 7. Cluster size distribution from the IKA among all the three damage regimes. Distribution of interstitial clusters at a) 0.0048 dpa, b) 0.0430 dpa, c) 0.1068 dpa and vacancy clusters at d) 0.0430 dpa, e)0.1068 dpa and f) ratio of the cluster count in the >41 and 2-4 cluster ranges (bins) as a function of damage(dpa).*

**Dislocation Formation, Structure and Evolution**

At higher damage, material microstructure evolves beyond point defects and clusters to form extended defect structures, including dislocation loops, network segments, and stacking-fault tetrahedra (SFTs). These extended defects can be expected to play a central role in determining mechanical response of the material and long-term radiation stability. To capture this stage of microstructural development, we analyze the nucleation, growth, and evolution of dislocation-based defects including SFTs across increasing damage levels. Our analysis uses the dislocation algorithm in OVITO which captures all possible dislocation types in the system: 1/3⟨111⟩ Frank loops, 1/6⟨112⟩ Shockley partial loops or chains, 1/6⟨110⟩ stair-rod loops, 1/2⟨110⟩ perfect dislocations, 1/3⟨100⟩ Hirth dislocations, and SFTs calculated from the stair-rod segments. To estimate the prevalence of SFTs, we use as a reference the idealized geometry of a perfect (non-truncated) tetrahedron, in which each SFT is bound by six sessile stair-rod dislocations, one along each tetrahedral edge. By counting the total number of stair-rod segments in the simulation cell and dividing by six, we obtain an approximate SFT population [23]. While not every stair-rod segment necessarily participates in a perfect tetrahedral assembly, making our SFT count somewhat of an approximation, this procedure gives a consistent, comparative metric for assessing relative SFT formation across damage states. The number of SFTs can also be approximated by filtering HCP-classified atoms; however, this approach may lead to an even greater overestimation because HCP coordination in FCC structures can additionally originate from coherent twin boundaries, Shockley partial dislocations, and highly distorted regions surrounding point-defect clusters

In Fig. 8a-g, we show 3D snapshots of the dislocation structures present at eight representative damage levels, including one from the initial damage regime (0.015 dpa), three from the second damage regime (0.035 dpa, 0.054 dpa, 0.074 dpa), and three from the third damage regime (0.94 dpa, 0.132 dpa and 0.161 dpa). In the low damage regime (Fig. 8a), largely isolated circular 1/3⟨111⟩ loops dominate, with few 1/6⟨112⟩ chains and 1/3⟨111⟩ loops. In the medium damage regime (Fig. 8b-d), many more dislocations begin to form and interconnect. By 0.054 dpa, as shown in Fig. 8c, 1/6⟨112⟩ partials weave through the volume, intersecting with an increasing number of 1/6⟨110⟩ junctions and 1/3⟨111⟩ loops. This intersection density grows further by 0.074 dpa in Fig. 8d, where a tangle of dislocations fills much of the simulation cell. In the final regime, perfect SFTs begin forming. This is evident at 0.094 dpa (Fig. 8e), where a few perfect SFT have already formed as shown in detail in Fig. 8h, along with an increasing number of 1/6⟨110⟩ partials (See the full supplementary video S1). At 0.132 dpa in Fig. 8f, an interconnected dislocation network densely fills the entire cell. Finally, at 0.164 dpa (Fig. 8g), 1/6⟨110⟩ segments dominate. Specifically, 1/6⟨112⟩ partials form an interconnected mesh, with the increasing number of 1/6⟨110⟩ segments and SFT, perhaps acting as rigid crosslinks at junctions, while there are a few large, isolated 1/3⟨111⟩ loops. Overall, there is a systematic increase in 1/6⟨110⟩ segments throughout the second and third regimes.

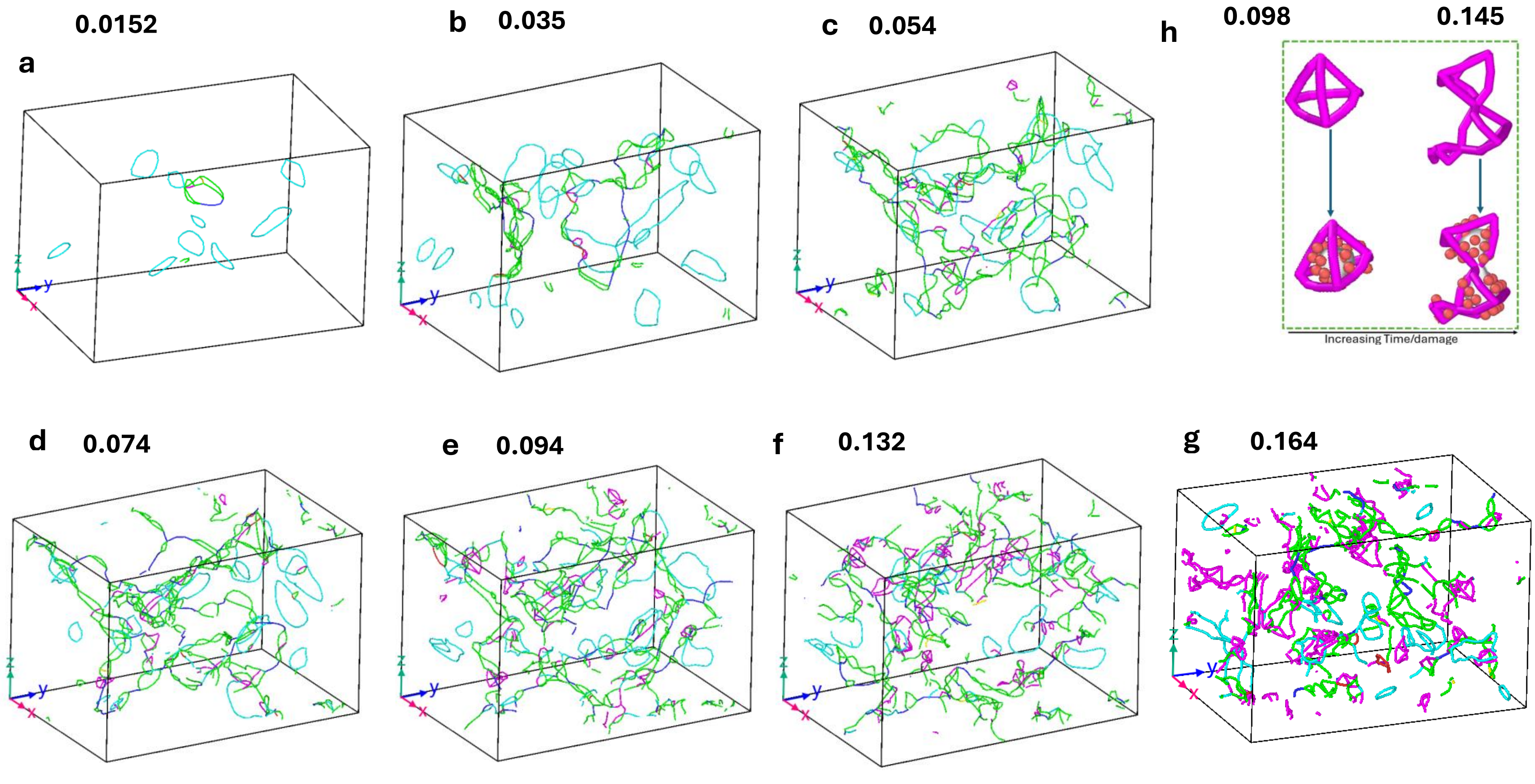

***Fig. 8.*** *Snapshots of the dislocation networks at different damage levels at a) 0.0152 dpa, b) 0.035 dpa, c) 0.054 dpa, d) 0.074 dpa, e) 0.094 dpa, f) 0.132 dpa, g) 0.164 dpa and h) showing a perfect individual SFT with core atoms and a coalesced or interacting SFTs showing slight changes in their orientation as damage increases. Blue, green, magenta, cyan, yellow, and red represent 1/2⟨110⟩ (perfect), 1/6⟨112⟩ (Shockley), 1/6⟨110⟩ (stair-rod), 1/3⟨111⟩ (Frank), 1/3⟨100⟩ (Hirth), and other dislocations, respectively, with the corresponding Burgers vector shown in Table I.*

**Table I.** *Dislocation types, including the color used in Figs. 8, 10-12, and 14, the dislocation type, and the corresponding Burgers vector.*

| Color | Type | Burgers Vector |
|---|---|---|
| Blue | Perfect | 1/2⟨110⟩ |
| Green | Shockley | 1/6⟨112⟩ |
| Magenta | Stair-rod | 1/6⟨110⟩ |
| Cyan | Frank | 1/3⟨111⟩ |
| Yellow | Hirth | 1/3⟨100⟩ |
| Red | Others | Varies |

We also carry out a more quantitative analysis of the dislocation evolution with increasing damage. Fig. 9a shows the total dislocation line length of individual dislocations. Below 0.01 dpa, there is virtually no dislocation activity in the system because the system has not yet absorbed enough defects to initiate a one-dimensional core misfit. Between 0.01 and 0.02 dpa, Frank 1/3⟨111⟩ loops appear and dominate. Above 0.02 dpa, 1/3⟨111⟩, Shockley 1/6⟨112⟩, and stair-rod 1/6⟨110⟩ partials begin to grow. By 0.04 dpa, the total dislocation line length exceeds 350 nm, corresponding to a density of $7.5\times10^{16}$ m$^{-2}$, as shown in Fig. 9c. Of this, Frank loops account for 140 nm ($3.0\times10^{16}$ m$^{-2}$), Shockley partials for 110 nm ($2.36\times10^{16}$ m$^{-2}$), and stair-rod partials for 60 nm ($1.29\times10^{16}$ m$^{-2}$) in Fig. 9a. With further damage, Shockley partials increase while Frank loops decrease, plateauing near 100 nm ($2.15\times10^{16}$ m$^{-2}$) by 0.1 dpa. As damage increases beyond 0.04 dpa, the total length of Shockley partials increases significantly, while the total length of Frank loops begins to decrease, plateauing at 100 nm by 0.1 dpa.

We also characterize the average dislocation segment length defined as the total individual dislocation line length divided by the number of segments (Fig. 9b). This provides a direct measure of dislocation connectivity and interaction. Dislocations composed of a small number of long segments tend to be structurally stable and participate weakly in dislocation–dislocation interactions, whereas dislocations consisting of many short segments are indicative of frequent junction formation, fragmentation, and network development (Fig. 9b, see also, the dislocation segment count in Fig. S8). Within the first damage regime (up to 0.03 dpa), the Frank segments exhibit a pronounced maximum length increase from 1 nm to 7 nm, before decreasing to approximately 3 nm throughout both the second and third damage regimes. By contrast, Shockley partials/loop segments maintain a nearly constant mean length of 1.5–2 nm throughout the simulation, regardless of the damage regime. The Frank loops are almost 100% pure edge while Shockley partials/loops are mixed dislocations loops as shown in Fig. S9. The almost circular morphology of the Frank loops throughout the simulations confirms that they are interstitial-typed loops, as was determined previously [24]. By their nature, it allows stress to readily concentrate along the faulted plane instead of the loop's edge. They continue to absorb interstitials; yet due to their sessile nature, they cannot glide away. Instead, they continue to grow into large sessile segments, driving their average segment length up to several nanometers. The Shockley partials, on the other hand, are mixed edge–screw and glissile by nature, which allows cross-slip and glide that both facilitate loop nucleation and limit continuous individual segment growth, yielding a much more uniform, shorter segment length (1 – 2 nm).

Fig. 9c shows the dislocation density in the three damage regimes. In the first half of the first damage regime, the overall dislocation density is almost zero because not enough damage has accumulated to initiate one-dimensional core misfits. In the second damage regime (from 0.03 – 0.09 dpa) the dislocation density increases rapidly up to about 1.50 x $10^{17}$ $m^{-2}$, indicative of significant damage and absorption of point defects which also explains why recombination decreases (Fig. 5a). It also indicates the formation and multiplication of dislocations as the material responds to irradiation. Going into the third regime (from 0.09 – 0.175 dpa), the density increase slows, suggesting that no significant dislocation nucleation is happening. The curve then begins to saturate, indicating a dynamic balance between dislocation nucleation and annihilation or rearrangement.

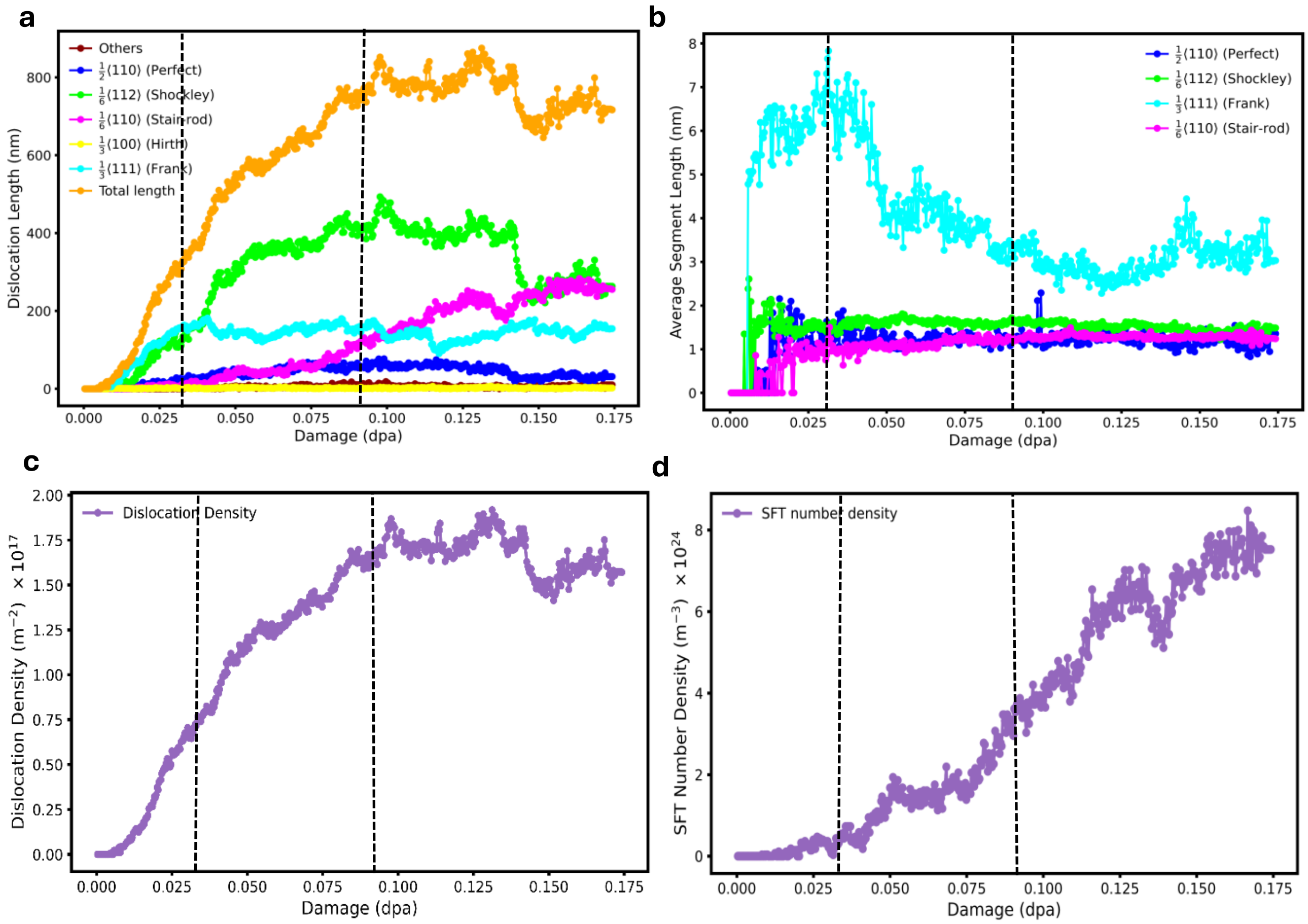

***Fig. 9.*** *Dislocation behavior with increasing damage, a) Total dislocation line length (nm) for different dislocation partials, b) Average dislocation segment length (nm), c) Dislocation density from the entire damage d) Number density of SFT arising from increasing damage. The black dash lines separate the three defect evolution regimes.*

Understanding the mechanism by which dislocations absorb or transform under irradiation is essential to unraveling how defect interactions drive the evolution of the dislocation network. To this end, we analyze sequences of atomic snapshots that capture the complete transformation of multiple dislocations. We present two distinct mechanisms. The first involves the absorption of two proximate Frank loops into multiple Shockley loops linked by stair-rod partials, as shown in Fig. 10a – f (See the full movie in supplementary video S2). At 0.0268 dpa, the loops remain distinct. By 0.0280 dpa, stair-rod segments nucleate at their ends, stitching the loops into a single "figure-eight" configuration. Between 0.0286 and 0.0290 dpa, approximately 70% of this structure transforms into Shockley loops connected by short segments. By 0.0293–0.0299 dpa, the transformation is complete, producing a polygonal network of nine Shockley loops linked by residual stair-rod and perfect 1/2⟨110⟩ segments as the remaining Frank character disappears. Thus, this transformation is a stair-rod–mediated "stitch-and-transform" mechanism driving loop coalescence and reconfiguration.

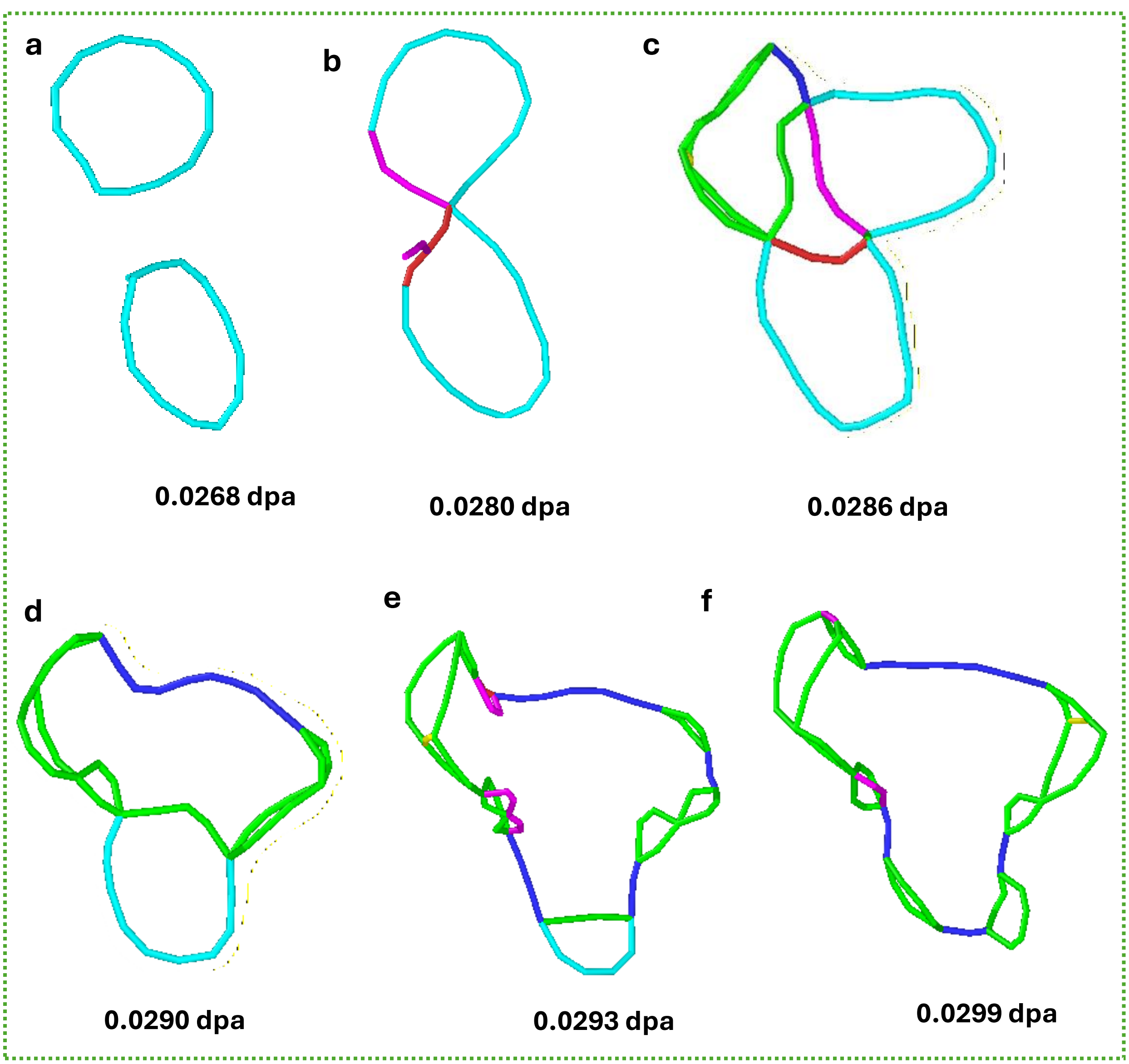

***Fig. 10.*** *Stitch-and-transform typed nucleation of Shockley loops. Atomic snapshots showing the evolution at a) 0.0268 dpa, b) 0.0280 dpa, c) 0.0286 dpa, d) 0.0290 dpa, e) 0.0293 dpa, f) 0.0299 dpa at increasing damage showing the first dislocation absorption/transformation mechanism. Two medium sized 1/3⟨111⟩ lying within a certain critical radius, partially transform into 1/6⟨110⟩ and eventually transform into 1/6⟨112⟩ chains. Movies are shown in supplementary video S2 (line coloring follows Table 1).*

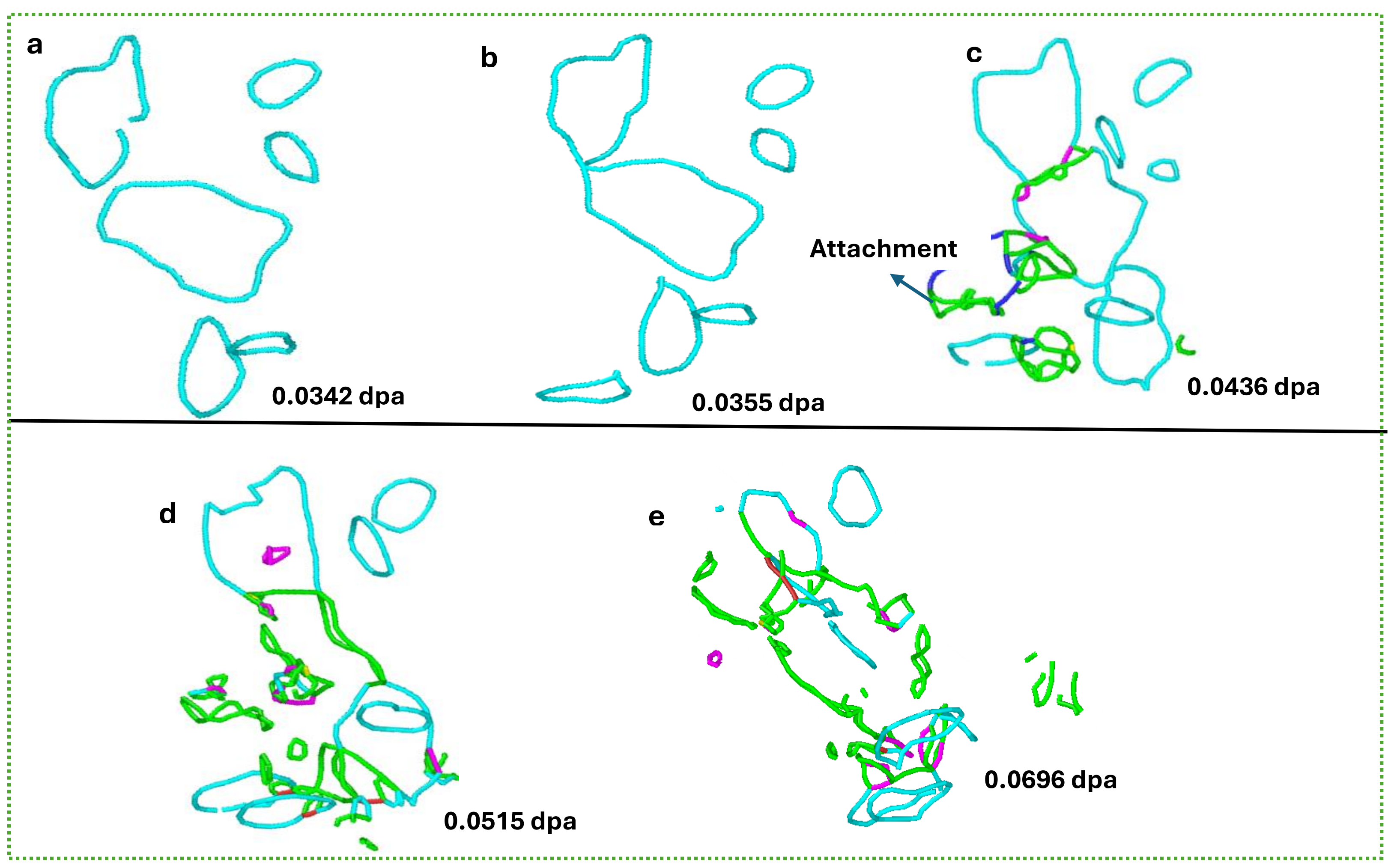

*Fig. 11. Dislocation loop coalescence-assisted partial transformation. Atomic snapshots taken at a) 0.0342 dpa, b) 0.0355 dpa, c) 0.0436 dpa, d) 0.0515 dpa, e) 0.0696 dpa showing a second dislocation absorption/transformation mechanism. Two individual 1/3⟨111⟩ loops grow into one large loop and transform into 1/6⟨112⟩ partials accelerated by secondary 1/6⟨112⟩ attachment and 1/6⟨110⟩ nucleation (segment coloring follows Table 1).*

The second mechanism involves the transformation of two large Frank loops surrounded by four smaller Frank loops into several interlocking rings of Shockley loops, as shown in Fig. 11(a – e); this transformation occurs within the second damage regime. By 0.0355 dpa, two large Frank loops merge and begin converting into Shockley loops; by 0.0436 dpa, a neighboring ring of Shockley loops attaches and triggers transformation from the opposite end; yet the process remains slower than stitch-and-transform. Around 0.0515 dpa, SFTs nucleate at the periphery and split into stair-rod segments that aid the conversion. By 0.0696 dpa, ~60% of Frank loops have dissociated into interlocking rings of Shockley loops. Overall, this is Frank-loop coalescence assisted by stair-rod loops and it requires more accumulated damage than the first mechanism.

It is well-known that SFTs arise from the collapse of stair-rod partials, and their formation and growth represent a key stage of microstructural evolution under irradiation [25]. Because SFTs is known to act as strong obstacles to dislocation motion, thus promoting pinning, hardening, and long-term defect retention, their presence is often closely linked to the radiation tolerance of FCC metals [15,26–29]. Understanding the nucleation and evolution of these structures with increasing damage, is thus essential for interpreting the high-damage behavior of Al. In Fig. 9d we plot the number of SFTs as a function of accumulated damage in our simulation cell. The number of SFTs remains small (around $0.5 \times 10^{24}$ $m^{-3}$) during the first damage regime and increases steadily to $4 \times 10^{24}$ $m^{-3}$ by the end of the second regime. During the third regime the number of SFTs increases further, reaching $8 \times 10^{24}$ $m^{-3}$ by the end of the simulation.

We now shift attention to understanding the details regarding the nature of transformations giving rise to SFTs throughout the damage process. Fig. 12a shows the nucleation of three stair-rod segments from Shockley partials at 0.1071 dpa (See the black circle). By 0.1075 dpa (Fig. 12b), the stair rods have grown, have more Shockley partials and are beginning to nucleate a truncated SFT (indicated by the arrow). By 0.1081 dpa, the average stair-rod segment length has increased and a clear SFT has nucleated from the pre-existing Shockley partials. By 0.1087 dpa, a significant portion of this dislocation network has transformed into stair-rod segments and one SFT has formed. Around 0.1094 dpa, four merged SFTs have developed but remain attached to the network. At 0.1123 dpa, the structure of four interconnected SFTs has matured and detached from the dislocation network as a standalone entity (Fig. 12h). Such a large, interconnected SFT is a result of SFT coalescence under continuous dislocation transformation.

Having established the evolution of the microstructure under increasing damage, we now examine its implications for macroscopic properties, particularly irradiation-induced hardening. We quantified the irradiation induced hardening contribution from different obstacles using the dispersed barrier hardening (DBH) model (Fig. 13). The DBH model estimates the yield stress increase Δσ as a function of the number density and size of obstacles (microstructure). The model application and description of relevant parameters and assumptions is discussed in the quantitative irradiation strengthening section, at the end of the paper. We looked at the contribution from five obstacles, $\Delta\sigma_{interstitials}$, $\Delta\sigma_{vacancies}$, $\Delta\sigma_{SFT}$, $\Delta\sigma_{Frank\ partials}$, $\Delta\sigma_{Shockley\ partials}$ and from there we calculated the total hardening ($\Delta\sigma_{total}$). Interstitial clusters dominate the early-stage hardening, accounting for nearly 90% of the total strengthening at low damage (< 0.02 dpa) and reaching 0.6 GPa. By 0.1 dpa, their contribution rises to 2.0 GPa, while vacancy clusters and SFTs contribute 1.0 GPa and 0.5 GPa, respectively. At higher doses, the contributions from interstitial and vacancy clusters approach saturation. In contrast, SFT-induced strengthening continues to increase, driven by the progressive formation of stair-rod dislocations from Shockley partial reactions, which sustain SFT nucleation and growth.

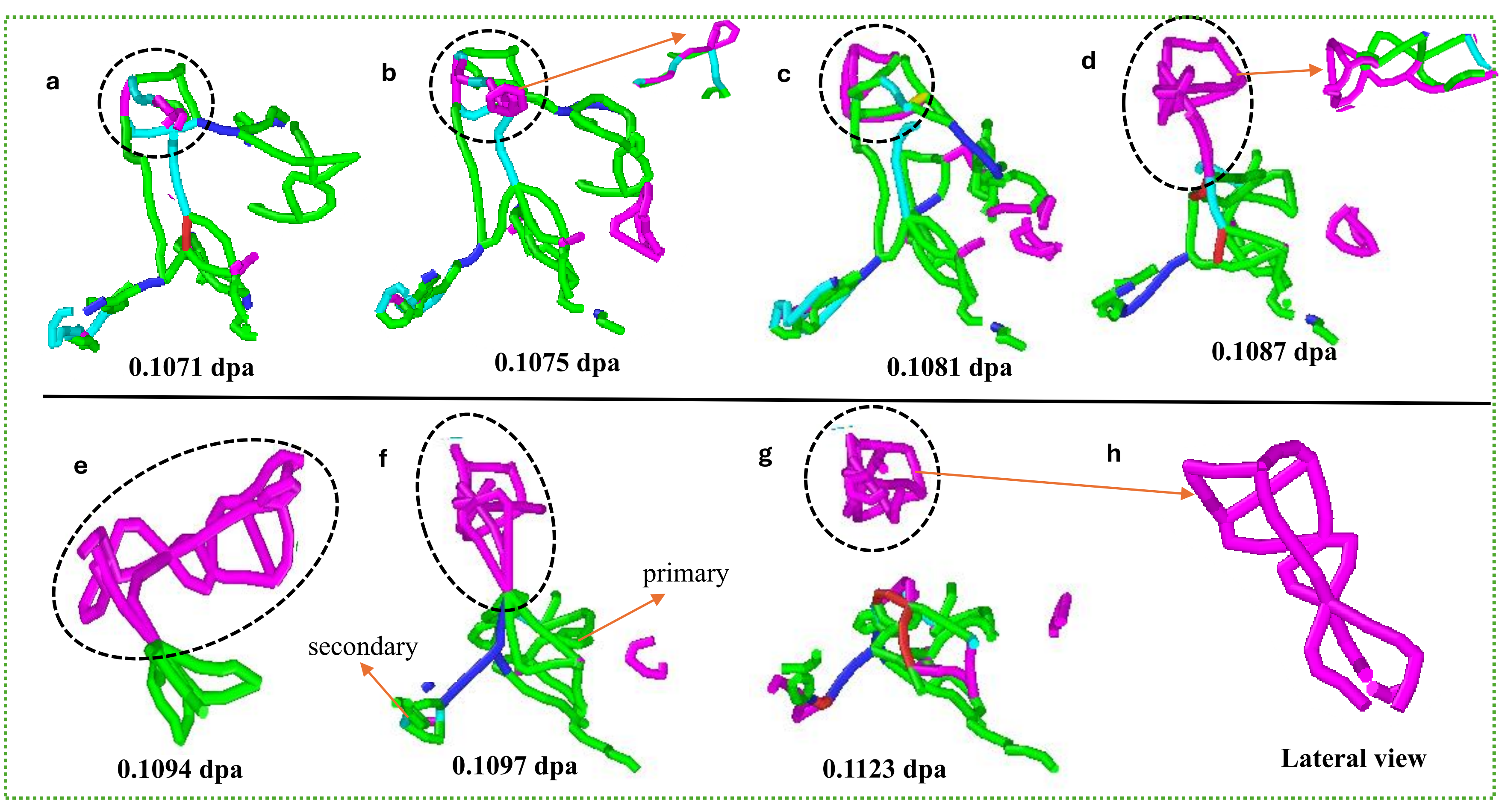

***Fig. 12***. *Mechanism of SFT nucleation via synergistic Shockley–Frank partial interactions. Nucleation and transformation of individual 1/6⟨110⟩ segments from 1/6⟨112⟩ Shockley partials into SFTs and their subsequent coalescence. The progressive nucleation is shown at a) 0.1071 dpa, b) 0.1075 dpa, c) 0.1081 dpa, d) 0.1087 dpa, e) 0.1094 dpa, f) 0.1097 dpa, g) 0.1123 dpa, h) extended lateral view of matured coalesced SFTs shown in g. The dashed black circles are for emphasis to guide the reader's eye*

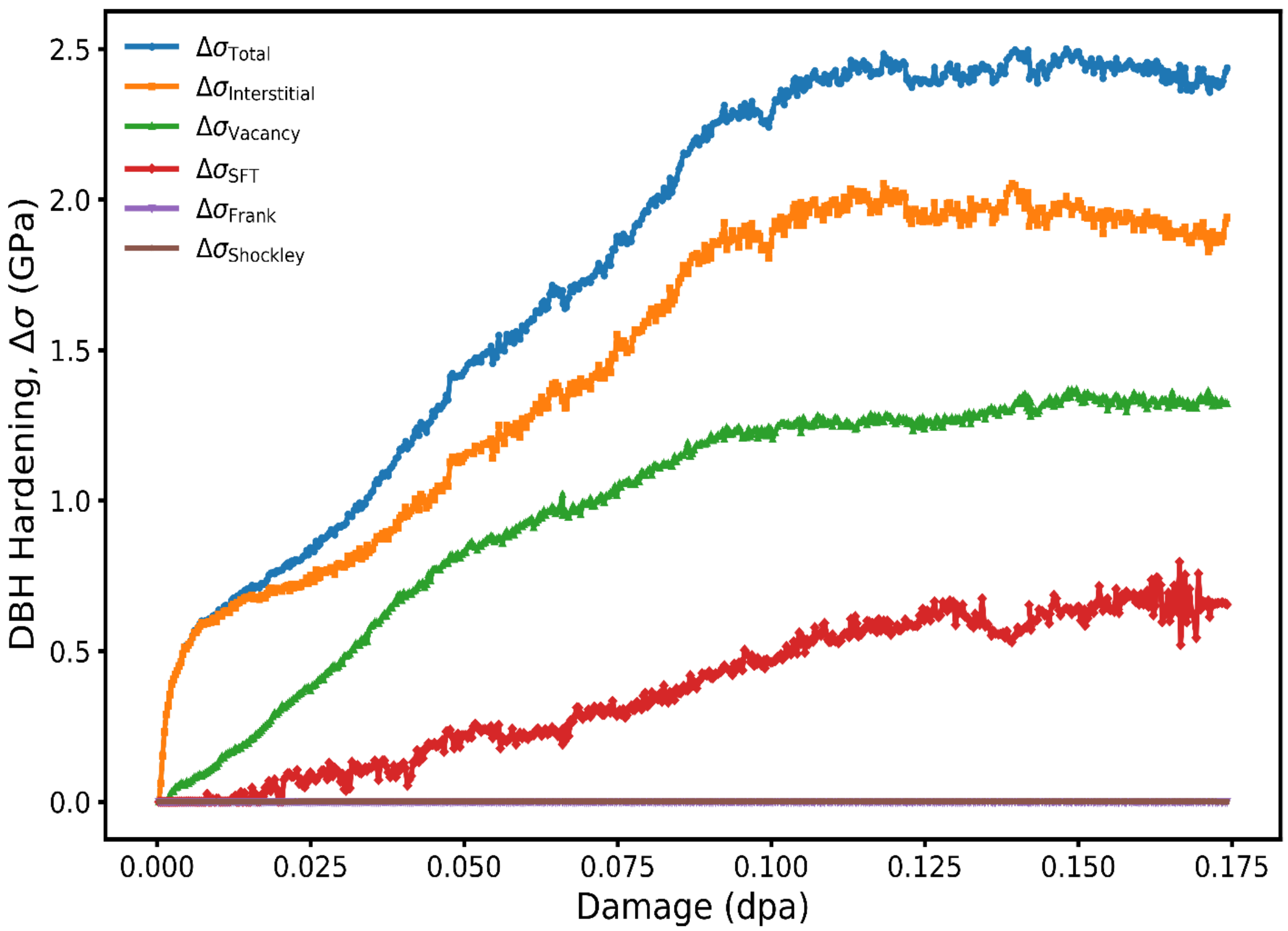

*Fig. 13. Hardening contribution from different obstacles. Irradiation induced hardening contribution from five different obstacles,* $\Delta\sigma_{interstitials}$ ,$\Delta\sigma_{vacancies}$ , $\Delta\sigma_{SFT}$ , $\Delta\sigma_{Frank\ partials,}$ $\Delta\sigma_{Shockley\ partials}$

## Discussion

The current study considers only single-crystal Al. We recognize that previous studies have confirmed that the presence of multiple grain boundaries may lead to significant differences due to grain boundary sink mediated behavior [30,31]. Extending it to polycrystalline materials and other crystal systems such as BCC is something we will consider in the future. Our result (Fig. 1) has shown that interstitial cluster growth in the CA simulations proceeds more gradually and only saturates and converges with the results of the IKA simulations at higher damage, consistent with prior observations [23,32,33]. We believe this is due to the differences in the defect generation. In Fig. 1c, the distinct evolution of the average cluster size at low-to-mid damage (< 0.045 dpa) reflects fundamental differences in defect generation and spatial correlations between IKA and CA. In IKA, rapid cluster coalescence arises from the stabilized, sequential insertion of Frenkel pair seeds, which minimizes stochastic defect–defect spacing and promotes early defect interaction. By contrast, in the CA the ballistic regime introduces strong spatial dispersion, distributing defects over larger distances, thereby delaying cluster coalescence. Beyond 0.03 dpa, some of the very large clusters in IKA begin to restructure or break into smaller, more stable configurations. This behavior is characteristic of Stage II–IV dynamic recovery, marked not only by a reduction in average cluster size but also by active dislocation reactions and annihilation [34,35]. Consequently, both methods show similar average cluster sizes beyond ~0.048 dpa.

The IKA reproduces the essential physics of cluster evolution even though the continuous introduction of fresh FPs in the IKA coupled with the absence of directional effects yields a higher overall defect concentration than the CA. In Fig. 3a–g, the cluster size distribution show that, despite minor quantitative differences at very low damage, the overall evolution of clusters in IKA and CA becomes remarkably similar as damage increases. The fraction of clustered atoms shown in Fig. 3h further confirms this point. Except for the smallest clusters (size 2–4) and the very largest (>41), the cluster-size distributions in the two methods become nearly identical at high damage. This suggests that once sufficient damage has developed, both IKA and CA follow similar mechanisms governing cluster nucleation, coalescence, and restructuring largely because the stored energies of the defects are similar at this point. The vacancy cluster density of $4.5\times10^{26}$ $m^{-3}$ at 0.12 dpa is comparable to the $2.4\times10^{23}$ $m^{-3}$ density at 0.05 strain recorded in copper [36]. This is ~3 orders of magnitude higher because IKA results in twice as high a defect concentration than the CA. These similarities support the conclusion that IKA reproduces the essential physics of cascade-driven cluster evolution, even though it bypasses the ballistic stage.

One potential criticism of the IKA simulation approach is that fixed-volume IKA simulations accumulate internal stress, which may influence defect mobility and clustering. To identify the potential impact of such stress, we compare the defect dynamics in a stressed system to a similar simulation where the stress generated by Frenkel pair addition was actively relieved by allowing the volume to relax. Allowing the volume to relax provides a control case that reveals whether the intrinsic stress biases microstructural evolution. Each stress-relaxation interval between consecutive IKA iterations was 35 ps for a cumulative time of 8.05 ns and up to 0.07 dpa. Because 35 ps is insufficient to fully relax long-wavelength stress, a residual hydrostatic stress of

~3 MPa remains in the cell. We will call this method the "relaxed" system and the other simulation where the stress is not relieved, the "stressed" system. The dynamics of interstitial evolution are shown in Fig. 14a-d and results of the total defects and vacancies are shown in Fig. S6, a-f.

Compared to the relaxed system in Fig. 14a, the intrinsically stressed system shows a somewhat lower number of interstitial clusters as damage accumulates. At the same time, the average cluster size in Fig. 8a and b is somewhat larger, particularly beyond 0.015, which is consistent with more efficient clustering and growth under stress. We attribute this to the FP-induced tensile stress (Fig. S7), which expands the lattice (i.e., increases interatomic spacing), and hence reduces the net defect migration barrier [37]. This stress-assisted mobility promotes defect migration and coalescence into bigger clusters as seen in Fig.14b. At low damage (≤ 0.02 dpa), the average cluster size is indistinguishable between both cases. Beyond this point, the stressed cell exhibits a sharp growth, peaking at 19 atoms/cluster by 0.035 dpa versus 13 for the relaxed system at the same damage. By 0.07 dpa both curves plateau, with the stressed cell stabilizing near 15 atoms/cluster and the relaxed cell near 12.5. The fraction of atoms in clusters normalized to the total defects sheds more light on the kinetics of clustering in both cases (Fig. 14c). Even though we see that the stressed cell has higher overall average cluster size especially at higher damage (Fig. 14b), the fraction of atoms incorporated into clusters overall is slightly higher in the relaxed cell up to 0.05 dpa, beyond which they are the same. This implies that the relaxed cell contains more clusters of smaller size whereas FP-induced stress favors coalescence/coarsening into fewer, larger clusters.

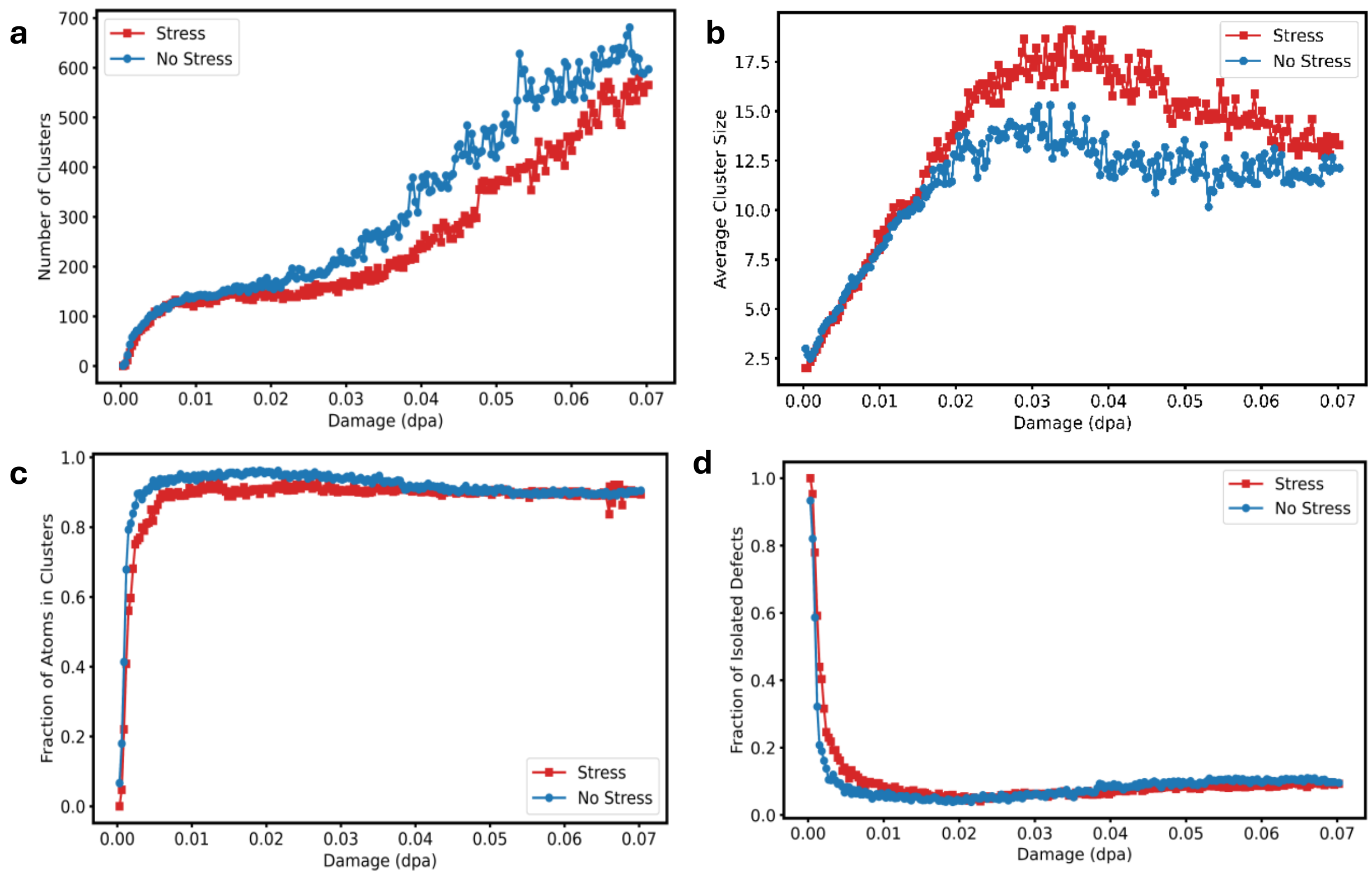

***Fig. 14.*** *Comparison of the impact of stressed and stress relaxation from the IKA simulations. a) Number of interstitial clusters, b) Average interstitial cluster size plot for the stressed and relaxed systems, **c)** Fraction of the atoms in clusters normalized to their corresponding total defect, d) Fraction of isolated point defects normalized to their corresponding total defects. Plot of the total defects for both the stressed and relaxed systems is shown in Fig. S6a.*

Finally, Fig. 14d examines the dynamics of isolated single point defects. The fraction of isolated defects decreases rapidly in both systems at very low damage and stabilizes around ~0.10 dpa. Up to 0.025 dpa, the relaxed system maintains a marginally lower fraction of isolated defects, implying slightly more efficient early clustering compared to the stressed system. Greater than 0.025 dpa, the two cases converge, consistent with the fraction of atoms in clusters plot shown in Fig. 14c, where both eventually saturate at similar values despite the stressed case producing somewhat larger clusters on average (Fig. 14b). Together, these trends underscore the complex role of FP-induced accumulated stress in governing defect dynamics. Accumulated stress suppresses the formation of small clusters while promoting coalescence of larger clusters and sustaining a slightly higher fraction of point defects. These insights are critical for understanding radiation damage evolution in stressed environments typical of nuclear materials, where intrinsic stress fields arise from pre-existing microstructural features or irradiation-induced lattice distortions. Although stress effects measurably influence the defect statistics in IKA simulations, their magnitude is insufficient to justify the substantial additional computational cost associated with including explicit stress relaxation in the IKA.

We next extend the discussion to the full IKA simulation up to 0.175 dpa, where the defect evolution can be divided into three distinct regimes with qualitatively different behavior: regime I, the early damage (0.001– 0.03 dpa), regime II, mid-damage (0.03 – 0.09 dpa), and regime III, high-damage (> 0.09 dpa). During regime I (0.005 – 0.04 dpa), the isolated vacancy fraction decreases as while as the SIA fraction (Fig. 5c), the vacancy cluster size and fraction do not change much, while the interstitial cluster size and fraction increase (Fig. 6), and the recombination fraction increases (Fig. 5a). In the low damage regime, the fast-moving SIAs find other interstitials to cluster and vacancies to recombine, but the slow-moving isolated vacancies have not yet begun to cluster. At slightly higher damage levels (~0.02 dpa), some of the interstitial clustering results in the formation of Frank loops and Shockley partials (Fig. 9a, b), increasing both total dislocation line length and average segment length. No significant SFT formation occurs during this early stage (Fig. 9d), underscoring that three-dimensional vacancy defects require time for the slow moving isolated vacancies to migrate and cluster. Nevertheless, even at these relatively low damage levels, incipient dislocation interactions are already developing, signaling the gradual transition toward extended defect–mediated microstructural evolution. Notably, the 'stitch-and-transform' absorption mechanism is active during this regime, activating at 0.0268 dpa and completing by the end of this regime. Examination of the full dislocation-evolution sequence (supplementary video S1) clarifies this behavior: Frank loops, being interstitial-type, grow rapidly by absorbing fast moving SIAs, increasing in size throughout the 0.001– 0.04 dpa range. However, once they reach a critical size, they become energetically unfavorable and are readily absorbed by the more stable Shockley loops, as illustrated in Fig. 10a–f. This explains both the sudden reduction in Frank loop length and in part, the onset of cluster-size saturation observed in this damage regime. Finally, the dominance of interstitial clustering in this regime directly translates to the observed mechanical response. As shown in Fig. 13, at low damage (< 0.025 dpa), hardening increases rapidly and is primarily governed by interstitial clusters, whose high mobility promotes early clustering and efficient obstruction of dislocation motion. This behavior is consistent with a recombination-dominated regime, where only the most stable interstitial clusters persist and act as effective obstacles.

In regime II (mid-damage regime), between 0.03 and 0.09 dpa, the isolated vacancy fraction continues to decrease slowly while the SIA fraction stabilizes (Fig. 5c), the vacancy cluster size and fraction and the interstitial cluster fraction increase while the interstitial cluster size decreases (Fig. 6), and overall annihilation fraction decreases (Fig. 5a) whilst survival increases (Fig. 5b). The isolated vacancies have had time to migrate and cluster and to interact with interstitial clusters, slowing their growth. Since isolated vacancies are tied up with clusters, recombination slows and defect accumulation (survival) increases significantly (Fig. 5b). The decrease in the isolated vacancy fraction continues but slows throughout this regime. Consistent with this shift, the total dislocation line length increases across most dislocation types over this damage range, signaling enhanced absorption and interaction with point defects; the notable exception is Frank partials, which exhibit a slight but systematic reduction, suggesting their absorption during dislocation restructuring. This reduction is attributed to the transformations identified in Fig. 10a–d and Fig. 11a–e, where Frank partials undergo Burger's vector changes from 1/3⟨111⟩ to 1/6⟨112⟩ Shockley loops. These transformations convert Frank partials into active sources for Shockley loop nucleation and multiplication [24,38], thereby redistributing dislocation character rather than eliminating dislocation content. This regime also has a pronounced increase in stair-rod formation (Fig. 9a). Stair-rod segments do not emerge at low damage but instead result from interactions among pre-existing dislocation loops and partials, a process that becomes increasingly probable once enough interstitial loops are present. Inspection of supplementary video S1 and Fig. 9a confirms that this transition initiates near 0.07 dpa, where loop–loop reactions and Burgers-vector transformations become active. As these reactions proceed, isolated vacancies and SIAs are absorbed, marking a transition from point-defect–production-dominated evolution to a dislocation-reaction–dominated regime. Concurrently, the SFT population increases, rising from $1.2 \times 10^{24}$ m$^{-3}$ at 0.03 dpa to $4 \times 10^{24}$ m$^{-3}$ at 0.09 dpa. The regime II (mid-damage regime) is therefore characterized by defect clusters reaching stable configurations through dislocation-mediated restructuring, while the continued introduction of FPs sustains elevated dislocation activity and accelerates microstructural complexity.

Dislocation transformation mechanisms similar to those we show in Figs. 10 and 11 were observe in equiatomic multicomponent Ni-based alloys at 0.024 dpa [23], closely matching the transformation and absorption events observed here at 0.027 dpa (Figs. 10a – d). The dislocation glide behavior identified in this regime is fully consistent with established slip-system physics [39]. Mobile Shockley partials preferentially glide on densely packed {111} planes, which are energetically favorable (Fig. 15). In contrast, Frank loops are sessile and energetically unstable; they readily dissociate into Shockley partials and subsequently into immobile stair-rod segments. The accumulation of these stair-rod segments at increasing damage progressively impedes Shockley partial motion, creating strong pinning points that promote their absorption and subsequent nucleation of SFTs which accelerate microstructural degradation (Fig. 15). Classical descriptions, such as the Silcox–Hirsch model [40], attribute SFT formation to the collapse of vacancy cluster into faulted Frank loops followed by dissociation into Shockley partials. Our simulations show similar mechanisms but also reveal an additional pathway in which SFTs form through the interaction and absorption of Shockley partials into stair-rod configurations. Thus, SFT formation proceeds via two concurrent mechanisms: vacancy collapse and dislocation-mediated evolution.

The reduction in recombination and the accompanying increase in defect accumulation directly drive the observed mechanical response in this regime, as the growing population of stable defect clusters increasingly impedes dislocation motion, leading to the continued rise in hardening (See Fig. 5a, b and Fig. 13). In this regime (~ 0.025 – 0.1 dpa), hardening continuous to increase primarily from three contributions thus interstitial clusters, vacancy clusters and a minor contribution from SFTs. This continued rise in total hardening reflects the transition to accumulation-dominated behavior, where defect production exceeds recombination and both defect populations co-evolve to control strengthening. The dislocation partials exhibited negligible contribution to hardening.

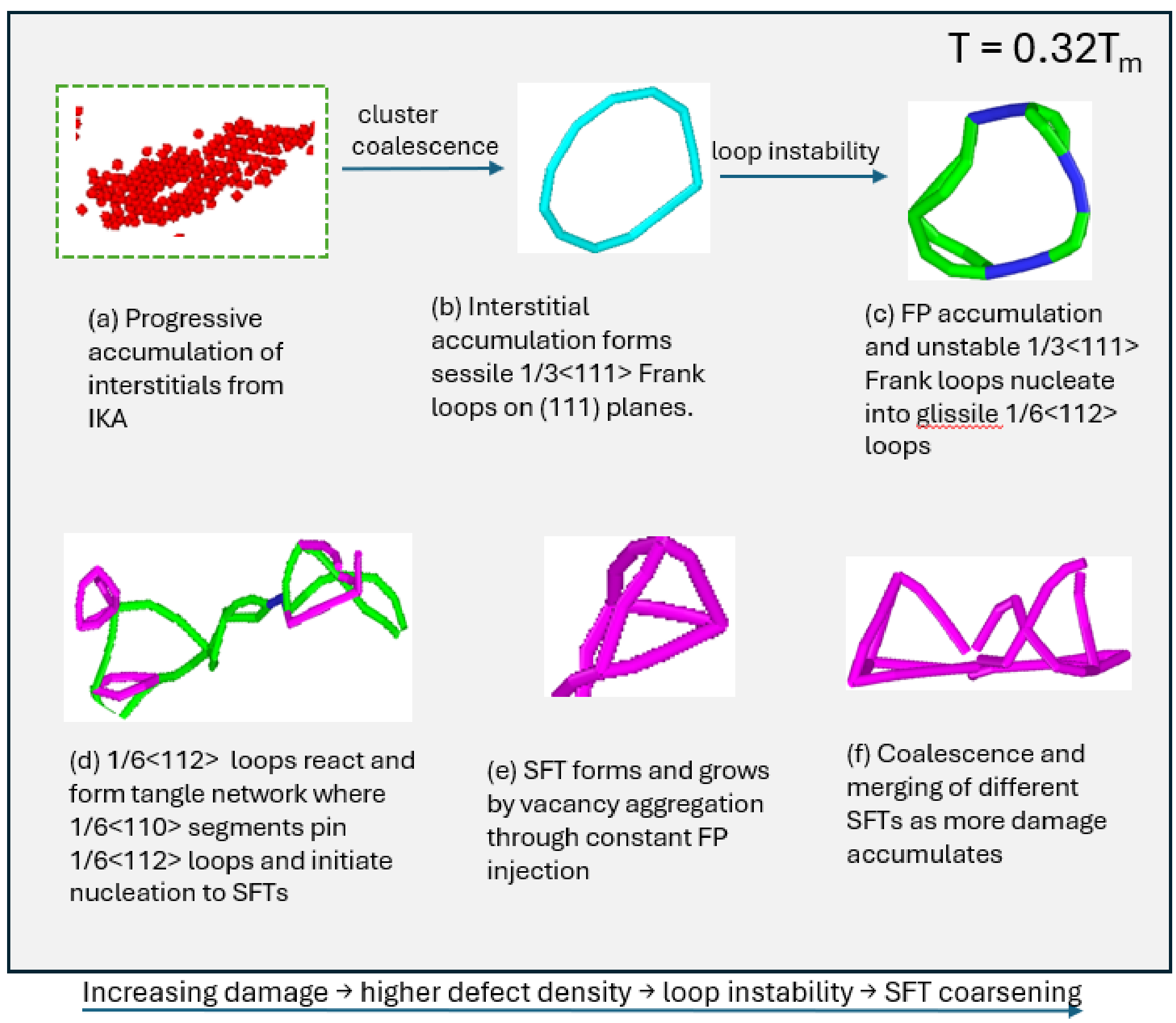

***Fig. 15.*** *Schematic illustration of the irradiation-induced dislocation transformation mechanistic sequence in Al with accumulated damage. Here we show the evolution of damage and the corresponding microstructural changes as damage accumulates, from the initial formation of clusters of different sizes through different loop nucleation to the formation of SFT and their subsequent coalescence*

In regime III (high damage regime), above 0.09 dpa, the fraction of isolated vacancies continues to decrease slowly, SIA, vacancy clusters, and interstitial clusters and the vacancy cluster size are all fairly unchanging (Fig. 4, Fig. 5c,  Fig. 6a, and Fig. S2). Also, both fraction of SIA and their and their concentrations are fairly unchanging (Fig. 5c and d). The average interstitial cluster size decreases slightly and then increase gradually, and the annihilation fraction increases rapidly. The increase in annihilation (Fig. 5a) is as a result of enhanced absorption at sink or higher sink strength due to enough dislocations units, not recombination. Another key microstructural process in this regime is the absorption of mobile Shockley partials by immobile stair-rod segments, leading to continued SFT formation (Figs. 11a–g). By this stage, Frank-to-Shockley transformations (Figs. 10 and 11) have largely saturated, and dislocation evolution is governed primarily by stair-rod–mediated reactions. This regime highlights the intrinsic coupling between point-defect and dislocation evolution, as defect clusters, whether interstitial or vacancy loops, are bound by partial dislocations and thus evolve through dislocation-mediated reactions. Consistent with this picture, all dislocation types except stair-rod partials remain largely stable without significant growth, whereas the stair-rod population continues to increase (Fig. 9a). As Shockley partials are progressively absorbed, mobile defect structures are converted into immobile configurations, effectively suppressing further interstitial cluster growth (Fig. 6a) and marking the transition to a dislocation-reaction-dominated steady-state regime.

The continued accumulation of stair-rod partials in this regime results in an increase in SFT nucleation and coalescence. As a result, pronounced SFT–SFT interactions emerge, driving the rapid increase in SFT density from $4 \times 10^{24}$ m$^{-3}$ at 0.1 dpa to $8 \times 10^{24}$ m$^{-3}$ at 0.16 dpa (Fig. 9d). As illustrated in Fig. 12a–h, mobile Shockley partials continue to be absorbed by stair-rod loops, promoting SFT formation, while neighboring SFTs interact and coalesce into larger, "giant" SFTs (See Fig. 12e and f). These merged SFTs progressively evolve toward a mature state, detaching from both primary and secondary dislocation networks and behaving as stable, self-contained defect structures (Fig. 12g and h). Tracking individual merged SFTs across successive damage increments reveals that most remain isolated rather than reattaching to dislocation lines until eventual annihilation, although a subset of truncated SFTs remains partially bound to surrounding dislocations. Consistent with this behavior, supplementary video S1 shows limited ongoing microstructural activity, with the dominant transformation being continued absorption of Shockley loops by stair-rod partials. At this regime (> 0.1 dpa), the hardening saturates (~2.4 – 2.5 GPa), consistent with the onset of sink-controlled or absorption dominated regime as clearly reflected in Fig. 5a and b. However, hardening contribution from SFTs continues to increase, although their overall impact remains lower than the clusters due to their lower number density thus resulting in overall saturation in total hardening.

At this stage, the total bulk dislocation density reaches $1.7 \times 10^{17}$ m$^{-2}$ at 0.15 dpa before declining to $1.4 \times 10^{17}$ m$^{-2}$ by 0.175 dpa, a level characteristic of severely deformed or highly irradiated crystalline materials. Dislocation densities of similar magnitude have been reported in both extreme deformation and irradiation studies. For example, Fan *et al.* [41] observed dislocation densities approaching $1 \times 10^{17}$ m$^{-2}$ in single-crystal Al and copper subjected to ultrahigh strain rates ($\dot{\varepsilon} \sim 10^{9}$ s$^{-1}$), accompanied by minimum yield stresses on the order of 10 GPa. Likewise, Chen *et al* [42] reported dislocation densities of $1.6 \times 10^{17}$ m$^{-2}$ in heavily irradiated Zr–Nb multilayers at 0.2 dpa, while Esfandiarpour *et al* [43] measured densities near $10^{16}$ m$^{-2}$in irradiated Inconel 617. These values far exceed the dislocation densities typical of cold-worked metals ($10^{10}$–$10^{12}$ m$^{-2}$) and fall within the regime associated with extreme plastic deformation and radiation-driven microstructural collapse ($10^{14}$ – $10^{17}$ m$^{-2}$) [44]. Collectively, this comparison demonstrates that the present approach drives the system into a damage state comparable to that of severely deformed crystals, reinforcing IKA capability to reproduce extreme irradiation-induced microstructural evolution.

In summary, we performed a systematic atomistic investigation of irradiation-induced microstructure evolution and degradation in single-crystal Al using the IKA benchmarked against conventional CA simulations. Quantitative comparison demonstrated that IKA reproduces the essential defect evolution kinetics and spatial defect topology observed in CA simulations, despite bypassing the explicit ballistic stage. At the same damage level (0.061 dpa), the interstitial pair-correlation functions from IKA and CA showed nearly identical spatial distributions with a Pearson correlation coefficient of $R = 0.998$ and RMSD = 0.0076, while equivalent-defect-fraction comparisons yielded $R = 0.984$. Although IKA promotes earlier interstitial clustering and larger early-stage cluster coarsening, both methods converge to similar interstitial and vacancy cluster behavior beyond 0.048 dpa, indicating that they capture the same underlying microstructural evolution mechanisms while enabling substantially accelerated access to higher damage.

Using the IKA, we extended irradiation damage accumulation up to 0.175 dpa and identified three distinct defect-evolution regimes governing microstructural evolution in Al. The total defect fraction increased rapidly to nearly 5% by 0.1 dpa before transitioning toward a dynamic steady state beyond 0.15 dpa. In the recombination-dominated regime (0.001– 0.03 dpa), the annihilation fraction increased to $\sim 29\%$, driven by rapid SIA migration and clustering. Interstitial clusters reached average sizes of 19 atoms by 0.03 dpa, nearly seven times larger than vacancy clusters at the same damage. In the accumulation-dominated regime (0.03 – 0.09 dpa), recombination decreased to nearly zero as vacancy clustering became increasingly active and dislocation restructuring accelerated. During this stage, the total dislocation density increased to $1.5 \times 10^{17}\ \mathrm{m}^{-2}$, accompanied by concurrent dislocation restructuring from $1/3\langle 111\rangle\ \ to\ 1/6\langle 112\rangle\ and\ \ to\ 1/6\langle 110\rangle$ transformations. In the high-damage sink-controlled regime (> 0.09 dpa), defect absorption by extended dislocation networks dominated the evolution, while SFT density increased from $4 \times 10^{24}\ \mathrm{m}^{-3}$ near 0.09 dpa to $8 \times 10^{24}\ \mathrm{m}^{-3}$ by 0.175 dpa through continued Shockley partial absorption and SFT coalescence.

Finally, the simulations further establish a direct mechanistic connection between microstructure evolution and hardening in Al. Interstitial loops dominated early hardening, contributing nearly 0.6 GPa below 0.02 dpa and increasing to 2.0 GPa by 0.1 dpa, while vacancy loops and SFTs contributed approximately 1.0 GPa and 0.5 GPa, respectively. Total irradiation hardening saturated near $2.4 - 2.5$ GPa at high damage as sink absorption and dislocation-mediated restructuring approached steady state. Overall, this work demonstrates that the IKA enables efficient access to high irradiation conditions while accurately capturing defect evolution physics from conventional cascade overlap simulations.

## Methods

The purpose of the IKA is to circumvent the high computational cost of explicitly resolving the ballistic regime of large numbers of displacement cascades, while establishing a direct correlation between the number of introduced Frenkel pairs and the incident ion energy. This accelerates damage accumulation, reaching higher damage levels than can be achieved with overlapping cascades. The IKA builds on several accelerated methods developed in the past. In this context, the conceptual idea of accelerated MD simulations can be traced back to the pioneering work of Limoge *et al.* [45], who studied radiation-induced amorphization through the successive introduction of Frenkel pairs or isolated interstitials at octahedral sites in metallic systems. Their study reported that irradiation damage evolution is rate-controlled, not solely dose-controlled, and that defect kinetics govern long-term structural evolution. This concept was later extended by Crocombette *et al.* [46] to study radiation-induced amorphization in pyrochlores using the Frenkel Pair Accumulation (FPA) method. Subsequent studies by Chartier *et al*. applied the FPA to investigate radiation resistance in pyrochlores [47] and irradiation-induced interstitial defect evolution in iron [48]. In the FPA method, an atom is displaced from its lattice site, leaving behind a vacancy and becoming an interstitial elsewhere. A vacancy–interstitial separation of 1.2 Å was enforced to reduce spontaneous overlap, followed by short relaxation periods of 0.5–2 ps. Goryaeva *et al*. used the FPA to predict the metastable state of interstitial dumbbells, also known as the A15 Frank-Kasper phase, a source of prismatic or faulted dislocation loops in Al [49]. Other studies have modified this approach by introducing single self-interstitial atoms (SIAs) sequentially [50].

While the extremely short evolution intervals between defect insertions in FPA-like methods can effectively mimic conditions where defect migration is strongly impeded (e.g., by impurities or solute trapping), they inherently underpredict local athermal Frenkel-pair recombination and preclude long-range thermally activated defect diffusion [51]. As a result, such approaches are not well suited for vacancy-type defects, whose migration typically requires long-range diffusion [50]. To address long-time defect kinetics, Aidhy *et al.* studied irradiation-induced point defect clustering in $UO_2$ [52] and MgO [53] using a single FP-loading approach, in which a fixed defect concentration (1 at.%) is inserted, followed by extended thermal evolution. Hendy and Ponga [54] developed a multiscale framework by coupling two-temperature molecular dynamics for the ballistic and thermal-spike stage with the Maximum-Entropy principle (MXE) to extend defect evolution to a mesoscale, allowing them to study both cascade damage and subsequent long-term defect migration and diffusion in nanocrystalline Nickel. It accounts for electronic energy dissipation and is useful for problems where electron-phonon coupling and post-cascade sink interactions are important. However, it is computationally demanding and coupling of two length scales could be significantly challenging.

More recently, Derlet and Dudarev [55] redefined the FPA concept by introducing the creation–relaxation algorithm (CRA) [55]. In the CRA method, an atom is randomly selected and inserted at a chosen location based on a probability distribution function and immediately followed by energy minimization. While this approach addresses data fluctuations encountered in the overlapping-cascade regime, due to immediate minimization without defect evolution it does not treat the high-energy component of defect production and does not permit long-range finite-temperature diffusion. The CRA was later modified to include spatial effects [56]. Vizoso et al. [57] combined explicit consecutive cascade simulations with an accelerated reduced-order atomistic cascade (ROAC) method to study radiation damage accumulation in gold and niobium nanostructures over a wide range of sizes and doses. In this method, ballistic defects are generated randomly by relocating atoms from a spherical core into a prescribed outer shell, thus introducing prescribed spatial correlations. A key strength of this approach is that it samples a recoil-energy spectrum rather than a single representative PKA energy. However, besides the computationally demanding nature of consecutive cascade simulations, the ROAC approximation caused inserted cascade regions to interact with free surfaces in the smallest nanowire system (10 nm), requiring explicit PKA simulations near the surface. Moreover, the ROAC model has been reported to predict larger average cluster sizes than MD cascades, particularly during the kinetic regime, with significantly higher cluster concentrations at the cell centroid (see original work [58])

Most existing accelerated methods discussed herein generally assume that added Frenkel pairs are uncorrelated with the incident ion energy, resulting in arbitrary number of inserted FPs. Consequently, direct comparison with experimental irradiation studies is challenging, as the absolute damage produced by accelerated MD methods cannot be quantitatively related to experiments due to the absence of a defined incident ion energy [19]. The IKA is modeled on the premise that cascade evolution consists of ballistic and kinetic stages [17], with residual defects from the ballistic stage governing subsequent kinetic processes. To bypass the costly ballistic stage, we use FP seeds calibrated from independent single-cascade simulations that account for athermal recombination corrected displacements per atom(arc-dpa) to correct overestimates from the traditional Norgett–Robinson–Torrens-dpa (NRT-dpa). Each inserted FP seed therefore represents one explicit cascade event corresponding to a well-defined incident ion energy, enabling direct comparison with experimental irradiation conditions while simultaneously accelerating damage accumulation. This makes IKA computationally efficient and yields comparable damage levels with defect morphologies and topologies similar to the traditional cascade-overlap simulations. The rest of this section focuses on describing the setup for the IKA methodology. However, to provide appropriate context, we first briefly discuss the more familiar methodology for the single and multiple cascades that are applied in this work to benchmark the IKA's accuracy.

**The Cascade Simulation:**

All MD simulations in this work, including both single and multiple collision cascade-overlap simulations, are performed using the LAMMPS code [59]. A PKA with energy of 5 keV, corresponding to a 50 keV helium ion incident energy as calculated by the SRIM code [60–62], is used. The PKA energy is estimated according to the guidelines provided by Stoller et al [63]. We construct the crystal with crystallographic orientations of [021], [0$\bar{1}$2], and [100] aligned along the x, y, and z axes, respectively, to mimic the crystallographic orientation used in our ongoing bicrystal simulations. The simulation domain measures 14.5 nm × 21.7 nm × 14.5 nm and contains 276,480 Al atoms. Based on previous studies that observed SFT formation and evolution in a significantly smaller systems, the present simulation cell is sufficiently large enough to capture the formation and evolution of SFTs [27]. Similarly, although dislocation loops can in principle interact with their periodic images under periodic boundary conditions, the present cell size is sufficiently large to minimize such interactions such that they do not significantly influence the results. To identify the most suitable interatomic potential, we calculate the root mean error between the predicted stacking fault energies and elastic constants for 32 candidate potentials documented in OPENKIM [64] and compare these results with experimental data reported by Noonon and Davis [65] and Kamm and Alers [66]. Through this analysis, we select an embedded atom method (EAM) potential developed by Mishin [67], as its elastic constants and stacking fault energies exhibit the lowest error with experimental values. This potential is well established and recognized for its accurate prediction of point and planar defects in FCC metals [68–70]. The structure is initially equilibrated at 300 K and zero pressure for 50 ps using the NPT ensemble, allowing the simulation box dimensions to fully relax. This is followed by an additional 450 ps of equilibration under the NVT ensemble, bringing the total equilibration time to 500 ps. This is sufficient to eliminate any residual stress in the system.

In the cascade simulation, we start with a well relaxed structure and randomly selected atom at the bottom center and give the atom velocity along the y direction equivalent to 5 keV of kinetic energy, as shown in Fig. 16. To capture short-range high energy interactions, we couple the EMA potential with the ZBL screened nuclear repulsion potential to screen high-energy collisions. The ZBL potential is applied at interatomic distances at $\leq 0.7$ Å. The structure is then minimized briefly using the conjugate gradient algorithm at zero pressure before the next cascade. The Langevin thermostat is used to control the temperature by applying frictional damping and random forces to atoms found within a 4 Å layer at all boundaries, which is thick enough to bring the system temperature back to the desired 300K.

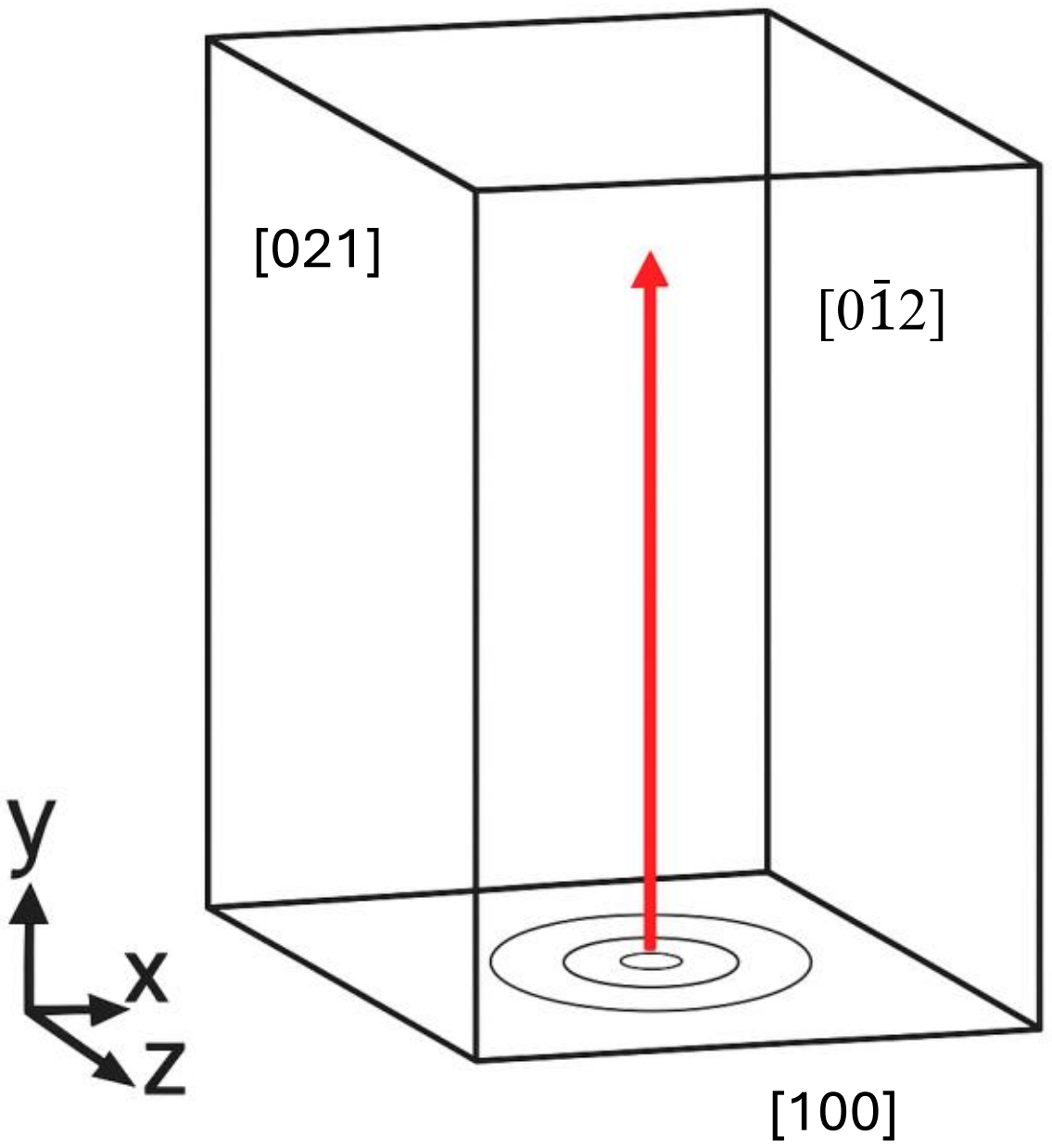

***Fig. 16.*** *Schematic of the PKA simulation box with x, y, and z directions corresponding to the [100], [021], and [0$\bar{1}$2] crystallographic directions. The bottom circles show the PKA region and the red arrow shows the direction of movement for the PKA. The IKA simulation cell has the same setup except that it does not involve the PKA region and direction defined here*

An adaptive time-step control is applied to ensure limited atomic displacements per step and to maintain numerical stability. Specifically, the dynamic timestep is set from $1.0 \times 10^{-6}$ ps, to $1.0 \times 10^{-3}$ ps with a maximum atom displacement threshold of 0.05 Å for the first 1-2 ps; thereafter, a timestep of $1.0 \times 10^{-3}$ ps is applied in the microcanonical ensemble (NVE). This prevents any unphysical atom overlaps or instabilities that can arise from large atomic velocities immediately after the PKA initiation. Periodic boundary conditions are applied in all three directions to mimic in infinitely large system. Finally, each cascade is allowed a maximum time of 35 ps to evolve before the next overlap.

**Iterative Kinetic Approach (IKA)**

For the IKA simulations, we use the same equilibrated structure with the same box dimensions as used in the cascade-overlap simulations described earlier. To calibrate the IKA simulation, we estimate the number of stable Frenkel pairs from the single cascade using the athermal recombination-corrected dpa (arc-dpa) using the equation

$$N_d = \xi \times N_{NRT}, \tag{2}$$

where ξ is the defect production efficiency factor, defined as the ratio of the actual number of stable defects to that predicted by the NRT model [4]. The efficiency factor ξ is given as

$$\xi(E_a) = \frac{1-C}{\left(2E_d/0.8\right)^b} E_a^b + c, \tag{3}$$

where $E_a$ and $E_d$ are the damage energy and the threshold displacement energies, respectively. We use a value of 24 eV for $E_d$ which is a typical threshold displacement energy for metals [71]. The parameters $b$ and $c$ are constants with values $b$ = -0.453 and $c$ = 0.481 from the works of Yin *et al* [72]. The arc-dpa estimated from the single cascade simulation resulted in 45 FPs. This number is therefore used as the FP "seed" in the IKA simulations. In each IKA iteration, 45 vacancies randomly located in the simulation cell are created and 45 interstitials are inserted at different random positions, representing the defect yield of one 5 keV cascade. After each FP insertion, the system is minimized to reach local energy minima and then evolved for 35 ps using two thermostatted regions, a 4 Å outer boundary layer, which is coupled to a Langevin thermostat to act as a heat sink, and an inner region, under a Nose–Hoover thermostat (NVT). This setup mimics the thermal response of cascade simulations by allowing localized overheating inside the cell while removing excess heat at the boundaries. Each iteration mimics a single cascade. Finally, to prevent unphysical atomic overlap, a 1.5 Å initial separation between inserted interstitial atoms and any other atoms was found ideal. This is slightly larger than the 1.2 Å separation commonly used in other accelerated-damage methods [51]. Finally, we estimate the cumulative damage displacement per atom (dpa) produced using the Norgett-Robinson-Torrens (NRT) equation but slightly modified to account for cumulative damage effect [32]. It is given as

$$\text{Damage (dpa)} = \frac{0.8T_d}{2E_d N} N_{IKA} \tag{4}$$

where N is the number of atoms in the simulation, $N_{IKA}$ is the number of iterations, and $T_d$ is the damage energy calculated from SRIM. We can also compute the effective damage/dose rate from the damage increment per IKA event using

$$\Delta D = \frac{N_{FP}}{N_{atoms}} \tag{5}$$

Where, ΔD is the damage increment per each IKA event, $N_{FP}$ is the number of inserted Frenkel pairs per IKA event, $N_{atoms}$ is the total atoms in the simulation cell. Therefore,

$$\Delta D = \frac{45}{276480} = 1.6276 \times 10^{-4}\ dpa \tag{6}$$

Since each IKA event was allowed a 35 ps evolution ($35\ \times 10^{-12}\ s$). The effective damage rate in the IKA simulations ($D_{IKA}$) is

$$D_{IKA} = \frac{1.6276 \times 10^{-4}\ dpa}{35\ \times 10^{-12}\ s} = 4.65 \times 10^{6}\ dpa\ s^{-1} \tag{7}$$

**Quantitative irradiation hardening**

The irradiation-induced hardening was estimated using the DBH model, which is widely applied to irradiation damage in metals [73,74]. For discrete three-dimensional defects such as clusters and SFTs, the DBH model expresses the yield strength increment as a function of obstacle size and number density:

$$\Delta\sigma_i = M\alpha_i \mu b\sqrt{N_i\ d_i} \tag{8}$$

where $\Delta\sigma_i$ is the strengthening increment (increase in yield stress) due to obstacle type $i$. $M$ is the Taylor factor, which is the shear stresses on a slip plane in a single crystal to the applied tensile stress required to activate slip in a polycrystal; for single-crystal Al, $M = 3.06$ [75]. $\alpha_i$ is the obstacle strength factor representing how effectively that defect impedes free dislocation motion. We assumed 0.1 – 0.5 for different obstacles strengths which is within close range with the values used in Arsenlis et al. [76]. $\mu$ is the shear modulus ($\mu = 26$ GPa for Al), and $b$ is the Burgers vector magnitude. For clusters, $b \approx 0.286$ nm, while for Shockley, Frank, and stair-rod dislocations (associated with SFTs), the corresponding magnitudes $0.165$, $0.234$, and $0.117$ nm were used, respectively. $N_i$ is the number density of obstacles ($m^{-3}$), and $d_i$ is their characteristic size ($m$). $i$ is the index denoting a specific defect type, e.g., SFTs, vacancy clusters, interstitial clusters, or dislocation loops. For vacancy and interstitial clusters, the mean diameter was estimated from the cluster volume using $V = n\Omega$, where $n$ is the cluster size and $\Omega$ is the atomic volume. Assuming spherical geometry, the equivalent diameter was obtained as

$$d = 2\left(\frac{3n\Omega}{4\pi}\right)^{1/3} \tag{9}$$

In contrast, for dislocation loops and line defects, the obstacle population is more appropriately described by the dislocation density $\rho_i$, defined as the total line length per unit volume ($m^{-2}$), which inherently incorporates both the number and spatial extent of the obstacles as was previously defined [76]. Their contribution to hardening is therefore expressed as

$$\Delta\sigma_i = M\alpha_i\mu b\sqrt{p_i} \quad (10)$$

where all parameters retain their definitions given above. For the dislocation contribution, only Shockley and Frank loops were considered as other dislocation types (e.g., perfect, Hirth, and others) contribute negligibly to the overall dislocation density. Although the total dislocation density could be used directly, this would lead to double counting because SFTs are composed of stair-rod dislocations, and their characteristic size is derived from the stair-rod edge length (i.e., the average segment length of stair-rod partials). To avoid this, the hardening contribution was decomposed by defect type. The total hardening is therefore taken as the root-sum-square superposition [77] of the individual contributions from each defect type in Eq. (11)

$$\Delta\sigma_{Total} = \sqrt{(\Delta\sigma_{interstitials})^2 + (\Delta\sigma_{vacancies})^2 + (\Delta\sigma_{SFTs})^2 + \left(\Delta\sigma_{Shockley}\right)^2 + (\Delta\sigma_{Frank})^2} \quad (11)$$

**Damage Analysis and Outlook**

All post-simulation data processing and analysis are performed using Python and a suite of algorithms and tools available within the Open Visualization Tool (OVITO) [78]. For point defect analysis, we employ the Wigner-Seitz cell algorithm [79], which offers the key advantage of a space-filling approach: each point in space is uniquely assigned to its nearest lattice site, eliminating the need for arbitrary cutoff distances as required in methods like Common Neighbor Analysis (CNA). A well equilibrated reference configuration is used for defect identification, and all reported data is based on averages from two independent simulations to ensure statistical reliability. We did not find the difference to be significantly different to justify the cost of conducting more than two independent simulations. Defects clusters are determined using a combination of the Wigner-Seitz cell algorithm [79] and the cluster analysis algorithm with a cutoff threshold of 3.2 Å.

Dislocations are characterized using the Dislocation Extraction Algorithm (DXA) [80]. DXA provides a robust, automated framework for identifying dislocation lines, dislocation cores, and readily classifying and outputting their corresponding Burgers vectors and types. In addition to the quantitative analysis of damage evolution presented in the sections below, the Supplementary

Material (SM) includes movies illustrating the evolution of dislocation networks and their transformations throughout the simulations.

In the following section, we discuss the limitations of the current study. In the current IKA implementation, localized spatial correlations arising from thermal spikes and subcascade formation are not explicitly considered. The PKA energy examined here lies below the established subcascade threshold in metals (10 –20 keV) [4,81,82]. Additionally, another simplification of the present IKA framework is that each IKA step is treated as a single representative PKA event rather than an ensemble of cascade events from a recoil-energy spectrum, as adopted in previous studies [57,58]. Together with the 35 ps relaxation interval between events, this yields an effective damage rate on the order of ~$10^6$, which remains above experimental irradiation rates (~$10^{-5}$–$10^{-3}$ dpa/s) but is lower than several previously reported accelerated atomistic methods, including FPA, the creation–relaxation algorithm, and ROAC [55,58,83]. This limitation, however, is not unique to IKA, but reflects a broader challenge of atomistic radiation-damage simulations [57,84]. Such high damage rates, as employed in the IKA approach and other MD cascade-overlap simulations, generate defects much faster than thermally activated recovery processes can operate under experimental conditions. Consequently, defect supersaturation and early-stage clustering may be enhanced relative to experiments, while long-range diffusion-mediated recovery mechanisms are reduced. Although this may accelerate the absolute kinetics and alter defect densities compared to low-dose-rate irradiation experiments, the underlying defect evolution pathways are generally expected to be preserved. Furthermore, the time allowed for inter-event diffusion (i.e., between successive IKA steps) is tunable in this approach. The interval of 35 ps, is designed to mimic the timescale over which an individual cascade evolves and cools down, which helps alleviate unrealistic defect accumulation by allowing local recombination and short-range defect rearrangements between damage events. Researchers interested in long term diffusive kinetics should refer to multiscale models that incorporates diffusion kinetics [54] or single FP-loading methods [85]

The introduction of artificial FPs, rather than explicit MD cascade overlaps, results in a slightly lower per-atom energy, decreasing from ~ −0.03 eV at 0.01 dpa to ~ −0.01 eV at 0.07 dpa [see our earlier work [86]. This may promote the formation of larger average clusters through reduced average atom–atom spacing, consistent with Fig. 2c. However, as damage accumulates, the difference progressively diminishes as the cascades stabilize. We considered only single-crystal

Al in the present study under room temperature (considered as 300K in this study). Also, exploring the impact of annealing at different temperatures will be something worth considering in the future. Further details of the IKA methodology and implementation are provided in our previous work [86]

## Acknowledgements.

We want to thank Dr. Brandon Scott Battas for useful discussions and editorial contributions. This research was funded by the United State Department of Energy (DOE), Office of Basic Energy Science Grant number DE-SC0024249.

### CRediT authorship contribution statement

**ASI:** Writing – original draft, Writing – review & editing, Visualization, Validation, Software, Methodology, Investigation, Formal analysis, Data curation, Conceptualization. **SW:** Writing – review & editing**. VY:** Writing – review & editing. **AA:** Writing – review & editing**,** Funding acquisition **MRT:** Writing – review & editing**,** Supervision, Funding acquisition, Project Administration. **SRP:** Writing – review & editing, Supervision, Funding acquisition, Methodology, Project Administration. All authors reviewed and confirmed the manuscript.

### Declaration of Competing Interest

The authors declare no competing financial or non-financial interests.

### Data availability

All supporting data for this article are included in the supplementary materials (SM). All correspondence should be directed at Michael R Tonks and Simon R Phillpot

### Code Availability

Codes will be made available upon reasonable request.

## References

1. Gregory, B. L. & Gwyn, C. W. Radiation effects on semiconductor devices. *Proc. IEEE* **62**, 1264–1273 (1974).
2. Holmes-Siedle, A. Radiation damage in semiconductor materials. *Nature* **280**, 176–176 (1979).
3. Li, S. *et al.* Development and applications of aluminum alloys for aerospace industry. *J. Mater. Res. Technol.* **27**, 944–983 (2023).
4. Nordlund, K. *et al.* Primary radiation damage: A review of current understanding and models. *J. Nucl. Mater.* **512**, 450–479 (2018).
5. Hasa, M. H. A. *et al.* Review of the development and application of aluminum alloys in the nuclear industry. *Rev. Adv. Mater. Sci.* **64**, (2025).
6. Woolstenhulme, N. E. *et al.* Irradiation performance of a U-7Mo in Al-Si matrix dispersion full-size fuel plate assembly. *Nucl. Eng. Des.* **416**, 112761 (2024).
7. Kim, H.-J., Yim, J.-S., Tahk, Y.-W., Oh, J.-Y. & Kong, E.-H. Drop behaviors of a plate-type fuel assembly used in research reactor for a drop accident. *Prog. Nucl. Energy* **113**, 255–262 (2019).
8. Al Hasa, M. H. *et al.* Review of the development and application of aluminum alloys in the nuclear industry. *Rev. Adv. Mater. Sci.* **64**, 20250122 (2025).
9. Busby, J. T., Nanstad, R. K., Stoller, R. E., Feng, Z. & Naus, D. J. *Materials Degradation in Light Water Reactors: Life After 60,* ORNL/TM-2008/170, 938766 http://www.osti.gov/servlets/purl/938766-MT7kYw/ (2008) doi:10.2172/938766.
10. Shepelyev, O., Sekimura, N. & Abe, H. Diffusion and conversion of interstitial dumbbells in segregated ternary alloys under irradiation. *J. Nucl. Mater.* **329–333**, 1204–1207 (2004).
11. Verbiest, E. & Pattyn, H. Study of radiation damage in aluminum. *Phys. Rev. B* **25**, 5097–5121 (1982).
12. Silcox, J. & Whelan, M. J. Direct observations of the annealing of prismatic dislocation loops and of climb of dislocations in quenched aluminium. *Philos. Mag. J. Theor. Exp. Appl. Phys.* **5**, 1–23 (1960).
13. Zhu, H. *et al.* The formation and accumulation of radiation-induced defects and the role of lamellar interfaces in radiation damage of titanium aluminum alloy irradiated with Kr-ions at room temperature. *Acta Mater.* **195**, 654–667 (2020).
14. Zhang, X. *et al.* Radiation effects on microstructure and hardening behavior in high-entropy fiber strengthened Al-based alloy. *J. Alloys Compd.* **1034**, 181244 (2025).
15. Satoh, Y., Yoshiie, T., Mori, H. & Kiritani, M. Formation of stacking-fault tetrahedra in aluminum irradiated with high-energy particles at low-temperatures. *Phys. Rev. B* **69**, 094108 (2004).

16. Chen, Y. *et al.* Heavy ion irradiation effects on CrFeMnNi and AlCrFeMnNi high entropy alloys. *J. Nucl. Mater.* **574**, 154163 (2023).
17. Aidhy, D. S. *et al.* Point defect evolution in Ni, NiFe and NiCr alloys from atomistic simulations and irradiation experiments. *Acta Mater.* **99**, 69–76 (2015).
18. Goryaeva, A. M. *et al.* Compact A15 Frank-Kasper nano-phases at the origin of dislocation loops in face-centred cubic metals. *Nat. Commun.* **14**, 3003 (2023).
19. Da Fonseca, D. *et al.* Evidence of dislocation loop preferential nucleation in irradiated aluminum under stress. *Scr. Mater.* **233**, 115510 (2023).
20. Jourdan, T., Goryaeva, A. M. & Marinica, M.-C. Preferential Nucleation of Dislocation Loops under Stress Explained by A15 Frank-Kasper Nanophases in Aluminum. *Phys. Rev. Lett.* **132**, 226101 (2024).
21. Wang, R. *et al.* Molecular dynamics simulation of effects of Al on the evolution of displacement cascades in Al CoCrFeNi high entropy alloys. *J. Nucl. Mater.* **577**, 154342 (2023).
22. Tawara, T., Matsukawa, Y. & Kiritani, M. Defect structure of gold introduced by high-speed deformation. *Mater. Sci. Eng. A* **350**, 70–75 (2003).
23. Levo, E., Granberg, F., Fridlund, C., Nordlund, K. & Djurabekova, F. Radiation damage buildup and dislocation evolution in Ni and equiatomic multicomponent Ni-based alloys. *J. Nucl. Mater.* **490**, 323–332 (2017).
24. Ma, K. *et al.* Decoding the interstitial/vacancy nature of dislocation loops with their morphological fingerprints in face-centered cubic structure. *Sci. Adv.* **11**, eadq4070 (2025).
25. Levo, E., Granberg, F., Nordlund, K. & Djurabekova, F. Temperature effect on irradiation damage in equiatomic multi-component alloys. *Comput. Mater. Sci.* **197**, 110571 (2021).
26. Schäublin *, R., Yao ,Z., Baluc ,N. & and Victoria, M. Irradiation-induced stacking fault tetrahedra in fcc metals. *Philos. Mag.* **85**, 769–777 (2005).
27. Aidhy, D. S. *et al.* Formation and growth of stacking fault tetrahedra in Ni via vacancy aggregation mechanism. *Scr. Mater.* **114**, 137–141 (2016).
28. Uberuaga, B., Hoagland, R., Voter, A. & Valone, S. Direct Transformation of Vacancy Voids to Stacking Fault Tetrahedra. *Phys. Rev. Lett.* **99**, 135501 (2007).
29. Zhang, L., Lu, C., Michal, G., Deng, G. & Tieu, K. The formation and destruction of stacking fault tetrahedron in fcc metals: A molecular dynamics study. *Scr. Mater.* **136**, 78–82 (2017).
30. Daghbouj, N. *et al.* Asymmetrical defect sink behavior of HCP/BCC Zr/Nb multilayer interfaces: Bubble-denuded zones at Nb layers. *Acta Mater.* **301**, 121579 (2025).
31. Daghbouj, N. *et al.* Nanoscale Strain Evolution and Grain Boundary-Mediated Defect Sink Behavior in Irradiated SiC: Insights from N-PED and DFT. *Acta Mater.* **303**, 121739 (2026).

32. Wei, G. *et al.* Revealing the critical role of vanadium in radiation damage of tungsten-based alloys. *Acta Mater.* **274**, 119991 (2024).
33. Granberg, F., Byggmästar, J. & Nordlund, K. Molecular dynamics simulations of high-dose damage production and defect evolution in tungsten. *J. Nucl. Mater.* **556**, 153158 (2021).
34. Nes, E., Marthinsen, K. & Brechet, Y. On the mechanisms of dynamic recovery. *Scr. Mater.* **47**, 607–611 (2002).
35. Szökefalvi-Nagy, A., Radnoczi, G. & Gaal, I. Stage IV recovery in deformed tungsten. *Mater. Sci. Eng.* **93**, 39–43 (1987).
36. Dai, Y. & Victoria, M. Defect structures in deformed f.c.c. metals. *Acta Mater.* **45**, 3495–3501 (1997).
37. Yang, W. & Wu, Y. Effects of annealing on the micro-internal stress induced by interstitial defects in aluminum crystal by molecular dynamics simulations. *AIP Adv.* **12**, 025226 (2022).
38. Anderson PM, Hirth JP & Lothe J. *Theory of Dislocations*. 3rd edn. Cambridge University Press (2017).
39. Mayer, A. E., Krasnikov, V. S. & Pogorelko, V. V. Dislocation nucleation in Al single crystal at shear parallel to (111) plane: Molecular dynamics simulations and nucleation theory with artificial neural networks. *Int. J. Plast.* **139**, 102953 (2021).
40. Silcox, J. & Hirsch, P. B. Direct observations of defects in quenched gold. *Philos. Mag.* **4**, 72–89 (1959).
41. Fan, H., Wang, Q., El-Awady, J. A., Raabe, D. & Zaiser, M. Strain rate dependency of dislocation plasticity. *Nat. Commun.* **12**, 1845 (2021).
42. Chen, E. Y., Deo, C. & Dingreville, R. Irradiation resistance of nanostructured interfaces in Zr–Nb metallic multilayers. *J. Mater. Res.* **34**, 2239–2251 (2019).
43. Esfandiarpour, A. *et al.* Characterizing the radiation-induced defect behavior in Inconel 617 through molecular dynamics simulations. *J. Nucl. Mater.* **616**, 156019 (2025).
44. Recovery, recrystallization & grain growth - NFPL139: Physics of Materials II.
45. Limoge, Y., Rahman, A., Hsieh, H. & Yip, S. Computer simulation studies of radiation induced amorphization. *J. Non-Cryst. Solids* **99**, 75–88 (1988).
46. Crocombette, J.-P., Chartier, A. & Weber, W. J. Atomistic simulation of amorphization thermokinetics in lanthanum pyrozirconate. *Appl. Phys. Lett.* **88**, 051912 (2006).
47. Chartier, A., Catillon, G. & Crocombette, J.-P. Key Role of the Cation Interstitial Structure in the Radiation Resistance of Pyrochlores. *Phys. Rev. Lett.* **102**, 155503 (2009).
48. Chartier, A. & Marinica, M.-C. Rearrangement of interstitial defects in alpha-Fe under extreme condition. *Acta Mater.* **180**, 141–148 (2019).
49. Goryaeva, A. M. *et al.* Compact A15 Frank-Kasper nano-phases at the origin of dislocation loops in face-centred cubic metals. *Nat. Commun.* **14**, 3003 (2023).

50. Monasterio, P. R., Yip, S. & Yildiz, B. Self-interstitial clusters in radiation damage accumulation: coupled molecular dynamics and metadynamics simulations. *Eur. Phys. J. B* **86**, 118 (2013).
51. Chartier, A. & Marinica, M.-C. Rearrangement of interstitial defects in alpha-Fe under extreme condition. *Acta Mater.* **180**, 141–148 (2019).
52. Aidhy, D. S., Millett, P. C., Desai, T., Wolf, D. & Phillpot, S. R. Kinetically evolving irradiation-induced point defect clusters in UO 2 by molecular dynamics simulation. *Phys. Rev. B* **80**, 104107 (2009).
53. Aidhy, D. S., Millett, P. C., Wolf, D., Phillpot, S. R. & Huang, H. Kinetically driven point-defect clustering in irradiated MgO by molecular-dynamics simulation. *Scr. Mater.* **60**, 691–694 (2009).
54. Hendy, M. & Ponga, M. A multiscale and multiphysics framework to simulate radiation damage in nano-crystalline materials. *J. Nucl. Mater.* **578**, 154347 (2023).
55. Derlet, P. M. Microscopic structure of a heavily irradiated material. *Phys. Rev. Mater.* **4**, (2020).
56. Mansouri, E. & Olsson, P. Modeling of irradiation-induced microstructure evolution in Fe: Impact of Frenkel pair distribution. *Comput. Mater. Sci.* **236**, 112852 (2024).
57. Vizoso, D. *et al.* Size-dependent radiation damage mechanisms in nanowires and nanoporous structures. *Acta Mater.* **215**, 117018 (2021).
58. Chen, E. Y., Deo, C. & Dingreville, R. Reduced-order atomistic cascade method for simulating radiation damage in metals. *J. Phys. Condens. Matter* **32**, 045402 (2019).
59. Thompson, A. P. *et al.* LAMMPS - a flexible simulation tool for particle-based materials modeling at the atomic, meso, and continuum scales. *Comput. Phys. Commun.* **271**, 108171 (2022).
60. Ziegler, J. F., Ziegler, M. D. & Biersack, J. P. SRIM – The stopping and range of ions in matter (2010). *Nucl. Instrum. Methods Phys. Res. Sect. B Beam Interact. Mater. At.* **268**, 1818–1823 (2010).
61. Daghbouj, N. *et al.* Defect landscape engineering suppresses helium damage in ceramics. *Commun. Mater.* **7**, 97 (2026).
62. Daghbouj, N. *et al.* Microstructure evolution of iron precipitates in (Fe, He)-irradiated 6H-SiC: A combined TEM and multiscale modeling. *J. Nucl. Mater.* **584**, 154543 (2023).
63. Stoller, R. E. *et al.* On the use of SRIM for computing radiation damage exposure. *Nucl. Instrum. Methods Phys. Res. Sect. B Beam Interact. Mater. At.* **310**, 75–80 (2013).
64. Tadmor, E. B., Elliott, R. S., Sethna, J. P., Miller, R. E. & Becker, C. A. The potential of atomistic simulations and the knowledgebase of interatomic models. *JOM* **63**, 17–17 (2011).
65. Noonan, J. R. & Davis, H. L. Truncation-induced multilayer relaxation of the A1(110) surface. *Phys. Rev. B* **29**, 4349–4355 (1984).

66. Kamm, G. N. & Alers, G. A. Low-Temperature Elastic Moduli of Aluminum. *J. Appl. Phys.* **35**, 327–330 (1964).
67. Mishin, Y. Atomistic modeling of the γ and γ'-phases of the Ni–Al system. *Acta Mater.* **52**, 1451–1467 (2004).
68. Chen, C., Qin, D., Wang, Y., Xu, F. & Song, J. Molecular dynamics simulations of interaction between a super edge dislocation and interstitial dislocation loops in irradiated L12Ni3Al. *J. Nucl. Mater.* **605**, 155541 (2025).
69. Nandipati, G., Senor, D. J., Casella, A. M. & Soulami, A. Molecular dynamics study of interstitial He clusters in nickel. *Nucl. Mater. Energy* **41**, 101733 (2024).
70. Anis, G., Hudson, T. & Brommer, P. Dislocation dynamics in Ni: Parameterizing dislocation trajectories from atomistic simulations. *Phys. Rev. Mater.* **8**, 123604 (2024).
71. Anderson, S., Khafizov, M. & Chernatynskiy, A. Threshold displacement energies and primary radiation damage in AlN from molecular dynamics simulations. *Nucl. Instrum. Methods Phys. Res. Sect. B Beam Interact. Mater. At.* **547**, 165228 (2024).
72. Yin, W., Zu, T. & Cao, L. Application of Athermal Recombination Corrected Dpa Model In Nuclear Data Processing Code Necp-Atlas. *EPJ Web Conf.* **247**, 09021 (2021).
73. Beck, M. S. & Iowa State Univ. of Science and Technology, A. (USA). Dispersed barrier hardening in irradiated metals. https://inis.iaea.org/records/ck1pw-t1277 (1978).
74. Chen, X. *et al.* Dispersed barrier hardening modeling on depth-distributed helium bubbles in iron-based alloys. *J. Nucl. Mater.* **606**, 155608 (2025).
75. Stoller, R. E. & Zinkle, S. J. On the relationship between uniaxial yield strength and resolved shear stress in polycrystalline materials. *J. Nucl. Mater.* **283–287**, 349–352 (2000).
76. Arsenlis, A., Rhee, M., Hommes, G., Cook, R. & Marian, J. A dislocation dynamics study of the transition from homogeneous to heterogeneous deformation in irradiated body-centered cubic iron. *Acta Mater.* **60**, 3748–3757 (2012).
77. Analysis of Obstacle Hardening Models Using Dislocation Dynamics: Application to Irradiation-Induced Defects | Metallurgical and Materials Transactions A | Springer Nature Link. https://link.springer.com/article/10.1007/s11661-015-2935-z.
78. Stukowski, A. Visualization and analysis of atomistic simulation data with OVITO–the Open Visualization Tool. *Model. Simul. Mater. Sci. Eng.* **18**, 015012 (2009).
79. Nordlund, K. *et al.* Defect production in collision cascades in elemental semiconductors and fcc metals. *Phys. Rev. B* **57**, 7556–7570 (1998).
80. Stukowski, A., Bulatov, V. V. & Arsenlis, A. Automated identification and indexing of dislocations in crystal interfaces. *Model. Simul. Mater. Sci. Eng.* **20**, 085007 (2012).
81. Muroga, T., Kitajima, K. & Ishino, S. The effect of recoil energy spectrum on cascade structures and defect production efficiencies. *J. Nucl. Mater.* **133–134**, 378–382 (1985).

82. Stoller, R. E. Primary Radiation Damage Formation. *Compr. Nucl. Mater.* https://doi.org/10.1016/B978-0-08-056033-5.00027-6 (2012) doi:10.1016/B978-0-08-056033-5.00027-6.
83. Crocombette, J.-P., Chartier, A. & Weber, W. J. Atomistic simulation of amorphization thermokinetics in lanthanum pyrozirconate. *Appl. Phys. Lett.* **88**, 051912 (2006).
84. Markelj, S. *et al.* Unveiling the radiation-induced defect production and damage evolution in tungsten using multi-energy Rutherford backscattering spectroscopy in channeling configuration. *Acta Mater.* **263**, 119499 (2024).
85. Aidhy, D. S., Millett, P. C., Desai, T., Wolf, D. & Phillpot, S. R. Kinetically evolving irradiation-induced point defect clusters in UO 2 by molecular dynamics simulation. *Phys. Rev. B* **80**, 104107 (2009).
86. Issaka, A. S. *et al.* Cascade-informed accelerated simulation of complex defect structures in irradiated aluminum using iterative kinetic approach. *Mater. Lett.* **414**, 140572 (2026).